\documentclass[useAMS,usenatbib,twocolumn]{mn2e}
\usepackage{amsmath}
\usepackage{amsfonts}
\usepackage{xcolor}
\usepackage[draft]{graphicx}

\makeatletter
\def\ref@jnl#1{{\rmfamily#1}}%
\newcommand\aj{\ref@jnl{AJ}}%
\newcommand\araa{\ref@jnl{ARA\&A}}%
\newcommand\apj{\ref@jnl{ApJ}}%
\newcommand\apjl{\ref@jnl{ApJ}}%
\newcommand\apjs{\ref@jnl{ApJS}}%
\newcommand\ao{\ref@jnl{Appl.~Opt.}}%
\newcommand\apss{\ref@jnl{Ap\&SS}}%
\newcommand\aap{\ref@jnl{A\&A}}%
\newcommand\aapr{\ref@jnl{A\&A~Rev.}}%
\newcommand\aaps{\ref@jnl{A\&AS}}%
\newcommand\azh{\ref@jnl{AZh}}%
\newcommand\baas{\ref@jnl{BAAS}}%
\newcommand\jrasc{\ref@jnl{JRASC}}%
\newcommand\memras{\ref@jnl{MmRAS}}%
\newcommand\mnras{\ref@jnl{MNRAS}}%
\newcommand\pra{\ref@jnl{Phys.~Rev.~A}}%
\newcommand\prb{\ref@jnl{Phys.~Rev.~B}}%
\newcommand\prc{\ref@jnl{Phys.~Rev.~C}}%
\newcommand\prd{\ref@jnl{Phys.~Rev.~D}}%
\newcommand\pre{\ref@jnl{Phys.~Rev.~E}}%
\newcommand\prl{\ref@jnl{Phys.~Rev.~Lett.}}%
\newcommand\pasp{\ref@jnl{PASP}}%
\newcommand\pasj{\ref@jnl{PASJ}}%
\newcommand\qjras{\ref@jnl{QJRAS}}%
\newcommand\skytel{\ref@jnl{S\&T}}%
\newcommand\solphys{\ref@jnl{Sol.~Phys.}}%
\newcommand\sovast{\ref@jnl{Soviet~Ast.}}%
\newcommand\ssr{\ref@jnl{Space~Sci.~Rev.}}%
\newcommand\zap{\ref@jnl{ZAp}}%
\newcommand\nat{\ref@jnl{Nature}}%
\newcommand\iaucirc{\ref@jnl{IAU~Circ.}}%
\newcommand\aplett{\ref@jnl{Astrophys.~Lett.}}%
\newcommand\apspr{\ref@jnl{Astrophys.~Space~Phys.~Res.}}%
\newcommand\bain{\ref@jnl{Bull.~Astron.~Inst.~Netherlands}}%
\newcommand\fcp{\ref@jnl{Fund.~Cosmic~Phys.}}%
\newcommand\gca{\ref@jnl{Geochim.~Cosmochim.~Acta}}%
\newcommand\grl{\ref@jnl{Geophys.~Res.~Lett.}}%
\newcommand\jcp{\ref@jnl{J.~Chem.~Phys.}}%
\newcommand\jgr{\ref@jnl{J.~Geophys.~Res.}}%
\newcommand\jqsrt{\ref@jnl{J.~Quant.~Spec.~Radiat.~Transf.}}%
\newcommand\memsai{\ref@jnl{Mem.~Soc.~Astron.~Italiana}}%
\newcommand\nphysa{\ref@jnl{Nucl.~Phys.~A}}%
\newcommand\physrep{\ref@jnl{Phys.~Rep.}}%
\newcommand\physscr{\ref@jnl{Phys.~Scr}}%
\newcommand\planss{\ref@jnl{Planet.~Space~Sci.}}%
\newcommand\procspie{\ref@jnl{Proc.~SPIE}}%

  \title[A needlet ILC analysis of WMAP 7-year temperature data]{A needlet ILC analysis of WMAP 7-year data: estimation of CMB temperature map and power spectrum}

  \author[Soumen Basak, Jacques Delabrouille] {Soumen
    Basak\thanks{E-mail: basak@apc.univ-paris7.fr}, Jacques
    Delabrouille\thanks{E-mail: delabrouille@apc.univ-paris7.fr}\\
Laboratoire APC, 10 rue Alice Domon et L\'eonie Duquet, 75205 Paris Cedex 13, France} \makeatletter

\begin{document}

\pagerange{\pageref{firstpage}--\pageref{lastpage}} \pubyear{2010}

\maketitle

\label{firstpage}


\begin{abstract}
 The WMAP satellite has provided high resolution, high signal to noise
 ratio maps of the sky in five main frequency bands ranging from $23$
 to $94$ GHz. These maps consist in noisy observations a mixture of
 Cosmic Microwave Background (CMB) anisotropies and of other
 astrophysical foreground emissions. We present a new
 foreground-cleaned CMB map, as well as a new estimation of the
 angular power spectrum of CMB temperature anisotropies, based on $7$
 years of observations of the sky by WMAP. The method used to extract
 the CMB signal is based on an implementation of minimum variance
 linear combination of WMAP channels and of external full-sky
 foreground maps, on a frame of spherical wavelets called
 \emph{needlets}. The use of spherical needlets makes possible
 localised filtering both in pixel space and harmonic space, so that
 the ILC weights are adjusted as a function of location on the sky and
 of angular scale.  Our CMB power spectrum estimate is computed using
 cross power spectra between CMB maps obtained from different
 individual years of observation. The CMB power spectrum is corrected
 for low-level biases originating from the ILC method and from
 foreground residual emissions, by making use of realistic simulations
 of the whole analysis pipeline.  Our error bars, compatible with
 those obtained by the WMAP collaboration, are obtained from the
 combination of two terms: the internal scatter of individual $C_\ell$
 in each $\ell$ bin, and a term originating from uncertainties in our
 correction for biases due to empirical correlations between CMB and
 foregrounds, as well as to residual foregrounds in the CMB maps.  Our
 power spectrum is essentially compatible, within error bars, with the
 result obtained by the WMAP collaboration, although it is
 systematically lower at the lowest multipoles, more than expected
 considering that the two estimates are based on the same original
 data.  Exhaustive investigations of the presence of a possible bias
 in our estimate fail to explain the difference.  Comparison with
 several other analyses confirm the existence of differences in the
 large scale CMB power, which are significant enough that until the
 origin of this discrepancy is understood, some caution is recommended
 in scientific work relying much on the exact value of the CMB power
 spectrum in the Sachs-Wolfe plateau.
\end{abstract}



\section{Introduction}

Cosmic Microwave Background (CMB) anisotropies provide a snapshot
of what the universe looked like at the moment of recombination. The
statistical properties of CMB fluctuations depend on the original
primordial perturbations from which they arose, as well as on the
subsequent evolution of the universe as a whole.  This makes the
precise measurement of CMB power spectra, both in temperature and
polarisation, a gold mine for understanding and describing the
universe from early times to now.

For cosmological models in which initial perturbations are Gaussian,
the information carried by CMB anisotropies is completely
characterised by the multivariate angular power spectrum of the
observations (i.e. the auto and cross power spectra of temperature and
polarisation maps).  The angular power spectrum of the CMB is
sensitive to the values of the different cosmological parameters which
primarily describe the fundamental properties of the universe, such as
its matter content, age, expansion history, global geometry and the
properties of the initial fluctuations that seeded the large scale
structure observable today.  The possibility to use this for
constraining the cosmological parameters hinges upon the existence of
acoustic oscillations in the primordial plasma before last scattering
of CMB photons, and on the ability of CMB experiments to disentangle
such primordial fluctuations from those induced by cosmic structures
at lower redshift.
This is why a key goal of the CMB community is to measure the true CMB
temperature anisotropies and polarisation in the sky. 

Since the discovery of the CMB by \citet{1965ApJ...142..419P},
tremendous efforts on the instrumental side have been made to improve
sensitivity and angular resolution of CMB experiments. Recently,
multi-frequency data for the complete sky have become available from 7
years of observation with the WMAP space mission
\citep{2010arXiv1001.4744J}. WMAP confirmed the leading theory in
cosmology, the $\Lambda$CDM (cold dark matter) model of a universe
governed by Einstein's theory of general relativity, and the evolution
of which is dominated by the impact of a cosmological constant
$\Lambda$, equivalent to a contribution, in the total energy density,
of vacuum energy, or more generally `dark energy'
\citep{2011ApJS..192...16L,2010arXiv1001.4538K}.

The WMAP data has already allowed a determination of several central
cosmological parameters with good accuracy. These determinations are
based on a number of simplicity assumptions, which need to be tested
by more accurate and extensive measurements. The Planck satellite of
the European Space Agency, next generation CMB space mission after
WMAP, surpasses its predecessor in resolution, sensitivity and
frequency coverage \citep{2010A&A...520A...1T,2011arXiv1101.2022P}, 
and will provide an
update of the present picture and of cosmological parameter estimates
in early 2013.

Determining the cosmological parameters with CMB experiments, however,
requires careful cleaning of the CMB maps from contamination by
galactic and extra-galactic foregrounds. In the low frequency
microwave regime (below about 100 GHz) the strongest contamination
comes from galactic synchrotron and free-free emission. At higher
frequencies, where synchrotron and free-free emissions are low, dust
emission dominates. In addition, extragalactic point-like sources are
a significant contaminant at high galactic latitude and on small
scale. Even when the brightest sources are identified and
subtracted-out or masked, their residual contribution may dominate
over the cosmic variance uncertainty on small angular scales, and bias
the measurement of the CMB power spectrum (see
\citet{2009LNP...665..159D} for a review on component separation in
CMB observations).

Most of the foreground emissions are strongly correlated between
frequency channels. Except for the very faint kinetic SZ signal, the
underlying frequency dependences of the emissions differ from that of
the CMB anisotropies. Exploiting this facts, a model-independent
method to remove foregrounds from the multi-frequency observations of
CMB, the so-called `Internal Linear Combination' (ILC), has been
proposed to extract the CMB signal from the multi-frequency data such
as that of WMAP or Planck (see, e.g., \citet{1996MNRAS.281.1297T}).
The idea behind the ILC method is to find the linear combination of
the available maps which has minimal variance while retaining unit
response to the CMB.  The main advantage of this foreground cleaning
method is that it does not require any assumption about the
foregrounds. Another advantage is that it is easy to implement, and
computationally fast. Finally, it can be extended to impose the
rejection of a particular foreground if needed
\citep{2011MNRAS.410.2481R}, or to extract the total combined emission
of several correlated foregrounds \citep{2011arXiv1103.1166R}.

There are, however, also drawbacks to the method.  As discussed by a
number of authors
\citep{2007ApJS..170..288H,2008PhRvD..78b3003S,2009LNP...665..159D,2009A&A...493..835D},
the component of interest (the CMB in our case) and foreground signals
must be uncorrelated for proper ILC performance. This can be, on
finite data sets, only approximately true, and empirical correlations
between the CMB and foregrounds generate a bias in the reconstructed
CMB.  In addition, as shown by \citet{2010MNRAS.401.1602D}, the ILC
method tends to amplify calibration errors, and reconstruct a CMB map
which can be largely under-calibrated, in particular for high
signal-to-noise ratio observations.  These sources of error must be
monitored carefully in the analysis of CMB maps obtained in this way.

In this paper, we address the problem of measuring as precisely as
possible the CMB temperature power spectrum from WMAP 7-year
observations, with close to minimal noise and
contamination by foreground emission. The paper is organised as
follows: In section \ref{sec:needlet-ilc} we describe the methodology
to estimate the CMB and its power spectrum using an ILC
on wavelet decompositions of sky maps observed at different frequencies. The implementation of this on
WMAP 7-year data is described in section \ref{sec:wmap-ilc} and
section \ref{sec:wmap-spec}. The results are discussed in section
\ref{sec:wmap-discuss}. We conclude in section
\ref{sec:conclusion}.


\section{The Needlet ILC}
\label{sec:needlet-ilc}

Component separation with an ILC method can straightforwardly be
performed in real space or in harmonic space. By this we mean that
different ILC weights can be computed in different regions of the sky,
or in different regions of harmonic space. This allows for variations
of the data covariance matrix in either space.

The ILC in harmonic space however does not take into account the fact
that noise is the dominant source of CMB measurement error at high
galactic latitude while foreground signals dominate at low galactic
latitude. On the other hand, the ILC in pixel space does not take into
account the fact that noise dominates at high angular frequency (small
scales) while foreground emission dominates on large scales. In order
to overcome this problem, we follow the approach of
\citet{2009A&A...493..835D} and implement the ILC on a frame of
spherical wavelets, called needlets. This special type of spherical
wavelets allows localised filtering in both pixel space and harmonic
space because they have compact support in the harmonic domain, while
still being very well localised in pixel domain
\citep{narcowich:petrushev:ward:2006,2008MNRAS.383..539M,guilloux:fay:cardoso:2008}.
Needlets have already been used in various analyses of WMAP data
besides component separation and power spectrum estimation, for
instance by \citet{2008PhRvD..78j3504P} to detect features in the CMB,
and by \cite{2009ApJ...701..369R} to put limits on the non-gaussianity
parameter $f_{\rm NL}$.

Various versions of the ILC on WMAP data have been implemented by a
number of authors so far
\citep{2003PhRvD..68l3523T,2004ApJ...612..633E,2007ApJS..170..288H,
  2007ApJ...660..959P,2008PhRvD..78b3003S,2009A&A...493..835D,2009PhRvD..79b3003K,2011BASI...39..163S}.
Alternate component separation on WMAP data has been performed using
independent component analysis methods such as SMICA
\citep{2005MNRAS.364.1185P}, CCA \citep{2007MNRAS.382.1791B}, and
FASTICA
\citep{2007MNRAS.374.1207M,2008MNRAS.389.1190B,2010MNRAS.402..207B}.
Specific analyses to extract foreground component emissions have
permitted to isolate the total foreground contamination
\citep{2011MNRAS.412..883G} or detect the thermal SZ emission of
galaxy clusters \citep{2008ApJ...675L..57A,2011A&A...525A.139M}.

\subsection{The data model}

A CMB space mission such as WMAP provides full-coverage, multi-frequency
anisotropy maps of the sky $T^{\text{OBS},c}(\hat{n})$ in $n_{c}$
different frequency bands (channels). The observed signal
$T^{\text{OBS},c}(\hat{n})$ in channel $c$ can be modelled as,
\begin{eqnarray}
T^{\text{OBS},c}(\hat{n})= \int_{\hat{n}^{\prime}} d\Omega_{\hat{n}^{\prime}}\,\,B^c(\hat{n},\hat{n}^{\prime})\,\,T^{\text{SIG},c}(\hat{n}^\prime) + T^{\text{N},c}(\hat{n}),
\label{eq:model}
\end{eqnarray}
where $T^{\text{SIG},c}(\hat{n})$ is the signal (sky) component, itself decomposed 
in the sum of CMB and foreground components,
\begin{eqnarray}
T^{\text{SIG},c}(\hat{n})=a^{c}\,T^{\text{CMB}}(\hat{n})+T^{\text{FG},c}(\hat{n}),
\end{eqnarray}
$a^{c}$ being
the CMB calibration coefficient for the channel $c$. Up to
calibration uncertainties, $a^c=1$
for all WMAP channels.  If, in addition to WMAP data, we use
extra ancillary data which serve as foreground templates to help
foreground subtraction, as done in the present work, the coefficients
$a^{c}$ vanish for such data sets.  

The beam function $B^{c}(\hat{n},\hat{n}^{\prime})$ represents the
smoothing of the signal due to the finite resolution of the observations.
The beam is assumed here to be
circularly symmetric, i.e.,
$B^{c}(\hat{n},\hat{n}^{\prime})$ depends only on the angle
$\theta=\cos^{-1}(\hat{n}.\hat{n}^{\prime})$ between the directions
$\hat{n}$ and $\hat{n}^{\prime}$. We may then expand this function in terms
of Legendre polynomials,
\begin{eqnarray}
B^c(\hat{n},\hat{n}^{\prime})=\sum_{l=0}^{\infty}\frac{2l+1}{4\pi}B_{l}^cP_{l}(\hat{n}.\hat{n}^{\prime}).
\end{eqnarray}
The term $T^{\text{N},c}(\hat{n})$ in equation \ref{eq:model} represents the detector noise in channel $c$. Unlike
the CMB and foreground components, instrumental noise is not affected
by the beam function.

Equation \ref{eq:model} can be recast, in the spherical harmonic representation, as:
\begin{eqnarray}
a_{l m}^{\text{OBS},c}=a^{c}\,B_l^c\,a_{l m}^{\text{CMB}}+B_l^c\,a_{l
  m}^{\text{FG},c}+a_{l m}^{\text{N},c}.
\label{map_harm_obs}
\end{eqnarray}

\subsection{Implementation of the needlet transform}

Considering that each channel observes the sky at a different
resolution, the maps are first convolved/deconvolved, in harmonic space, to the same resolution:
\begin{eqnarray}
a_{l m}^{c}=\frac{B_l}{B_l^c}\,\,a_{l m}^{\text{OBS},c}.
\label{map_harm}
\end{eqnarray}
Each of
these maps $a_{l m}^{c}$ is then decomposed into a set of filtered maps
$a_{l m}^{c,j}$ represented by the spherical harmonic coefficients,
\begin{eqnarray}
a_{l m}^{c,j}=h_{l}^{j}a_{l m}^{c}.
\end{eqnarray}
The filters $h_{l}^{j}$ is chosen in such a way that
\begin{eqnarray}
\sum_{j}\left(h_{l}^{j}\right)^{2}=1,
\end{eqnarray}
which permits us to directly reconstruct the original maps $a_{l
  m}^{c}$ from their filtered maps $a_{l m}^{c,j}$ using
the same set of filters. 
In terms of $h_{l}^{j}$, the spherical needlets are defined as,
\begin{eqnarray}
\Psi_{j k}(\hat n)=\sqrt{\lambda_{jk}}\sum _{l=0}^{l_{\max }}\sum_{m=-l}^l 
h_{l}^{j}\,Y_{l m}^{*}(\hat n)\,Y_{l m}(\hat\xi_{jk}),
\end{eqnarray}
where $\{\xi_{jk}\}$ denote a set of cubature points on the sphere,
corresponding to a given scale $j$. In practice, we identify
these points with the pixel centres in the HEALPix pixelisation scheme
\citep{2005ApJ...622..759G}. The cubature weights $\lambda_{jk}$ are
inversely proportional to the number $N_{j}$ of pixels used for the needlet decomposition at scale $j$, i.e.
$\lambda_{jk}=\frac{4\pi}{N_{j}}$.
The needlet coefficients for CMB temperature anisotropies
$T(\hat n)=\sum _{l=0}^{l_{\max }}\sum_{m=-l}^la_{lm}Y_{lm}(\hat n)$
are denoted as,
\begin{eqnarray}
\beta_{j k}&=&\int_{S^{2}}T(\hat n)\,\Psi_{j k}(\hat n)\,d\Omega_{\hat n}
\nonumber\\
&=&\sqrt{\lambda_{j k}} \sum _{l=0}^{l_{\max }}
\sum_{m=-l}^l h_l^j\,B_l\,a_{l m}\,\,Y_{l m}(\xi _{j
  k}).
\end{eqnarray}
The linearity of the needlet decomposition implies that the needlet 
coefficients $\beta_{j k}^{c}$ corresponding to
the filtered map obtained from the harmonic coefficients $a_{l
  m}^{c,j}$ are a linear combination of the needlet coefficients of
individual components and noise at HEALPix grid points $\xi_{j k}$:
\begin{eqnarray}
\beta_{j k}^{c}=a^{c}\,\beta_{j k}^{\text{CMB}}+\beta_{jk}^{\text{FG},c}+
\beta_{j k}^{\text{N},c}
\end{eqnarray}
where,
\begin{eqnarray}
\beta_{j k}^{\text{CMB}}&=&\sqrt{\lambda_{j k}} \sum _{l=0}^{l_{\max }}
\sum_{m=-l}^l h_l^j\,B_l\,a_{l m}^{\text{CMB}}\,\,Y_{l m}(\xi _{j
  k})\nonumber\\
\beta_{j k}^{\text{FG},c}&=&\sqrt{\lambda_{j k}} \sum _{l=0}^{l_{\max }}
\sum_{m=-l}^l h_l^j\,B_l\,a_{l m}^{\text{FG},c}\,\,Y_{l m}(\xi _{j
  k})\nonumber\\
\beta_{j k}^{\text{N},c}&=&\sqrt{\lambda_{j k}} \sum _{l=0}^{l_{\max }}
\sum_{m=-l}^l h_l^j\,\frac{B_l}{B_l^c}\,a_{l m}^{\text{N},c}\,\,Y_{l m}(\xi _{j
  k})
\end{eqnarray}

\subsection{Implementation of the needlet ILC}

The ILC estimate of needlet coefficients of the cleaned map is
obtained as a linearly weighted sum of the needlet coefficients
$\beta_{j k}^{c}$,
\begin{eqnarray}
\beta_{j k}^{\text{NILC}}=\sum_{c=1}^{n_c}\omega_{j k}^{c}\,\beta_{j k}^{c}
\end{eqnarray}
where $\omega_{j k}^{c}$ is the needlet weight for the scale $j$ and
the frequency channel $c$ at the pixel $k$. 
Under the assumption of de-correlation
between CMB and foregrounds, and between CMB and noise, the empirical
variance of the error is minimum when the empirical variance of the
ILC map itself is minimum. 
The condition for preserving the CMB
signal during the cleaning is encoded as the constraint:
\begin{eqnarray}
\sum_{c=1}^{n_{c}}a^{c}\omega_{j k}^{c}=1.
\label{constraints}
\end{eqnarray}
The resulting needlet ILC weights $\widehat{\omega}_{j k}^{c}$ that
minimise the variance of the reconstructed CMB, subject to the
constraint that the CMB is preserved, are expressed as:
\begin{eqnarray}
\widehat{\omega}_{j k}^{c}=\frac{\sum_{c^{\prime}=1}^{n_c}\left(\widehat{R}_{j
    k}^{-1}\right)^{c
    c^{\prime}}a^{c^{\prime}}}{\sum_{c=1}^{n_c}\sum_{c^{\prime}=1}^{n_c}a^{c}.
\left(\widehat{R}_{j
    k}^{-1}\right)^{c c^{\prime}}a^{c^{\prime}}}
    \label{eq:ilc}
\end{eqnarray}
The NILC estimate of the
cleaned CMB needlet coefficients is:
\begin{eqnarray}
\beta_{j k}^{\text{NILC}}=\beta_{j
  k}^{\text{CMB}}+\frac{\sum_{c=1}^{n_c}\sum_{c^{\prime}=1}^{n_c}\left(\beta_{j k}^{\text{FG},c}+
\beta_{j k}^{\text{N},c}\right)\left(\widehat{R}_{j k}^{-1}\right)^{c
    c^{\prime}}\,a^{c^{\prime}}}{\sum_{c=1}^{n_c}\sum_{c^{\prime}=1}^{n_c}a^{c}
\left(\widehat{R}_{j k}^{-1}\right)^{c c^{\prime}}a^{c^{\prime}}} \nonumber
\end{eqnarray}
where $\widehat{R}_{j k}^{c c^{\prime}}$ are empirical estimates of the elements of covariance
matrix $R_{j k}^{c c^{\prime}}=\left<\beta_{j k}^{c}\beta_{j
  k}^{c^{\prime}}\right>$ for scale $j$ at pixel $k$. Those estimates are obtained
each as an average of the product of the relevant computed needlet
coefficients over some space domain ${\cal D}_{k}$ centred at $k$. In practice, they are computed as
\begin{eqnarray}
\widehat{R}_{j k}^{c c^{\prime}}=\frac{1}{n_{k}}\sum_{k^{\prime}} w_j(k,k^\prime)
\beta_{j k}^{c}\beta_{j k}^{c^{\prime}},
\end{eqnarray}
where the weights $w_j(k,k^\prime)$ define the domain ${\cal D}_{k}$. 
A sensible choice is for instance
$w_j(k,k^\prime) = 1$ for $k^\prime$ closer to $k$ than some limit
angle, and $w_j(k,k^\prime) = 0$ elsewhere, or alternatively,
$w_j(k,k^\prime)$ shaped as a Gaussian beam of some given size that
depends on the scale $j$ (which is what we do here).

Finally, the NILC estimate of the cleaned CMB map can be reconstructed
from cleaned CMB needlet coefficients using the same set of filters
that was used to decompose the original maps into their needlet
coefficients. The NILC CMB temperature map is then
\begin{eqnarray}
T^{\text{NILC}}(\hat{n})&=&\sum_{l m} a_{l m}^{\text{NILC}}\,Y_{l
  m}(\hat{n})\nonumber\\ &=&\sum_{l m} \,\left(B_{l}\,a_{l
  m}^{\text{CMB}}+a_{l m}^{\text{RFG}}+a_{l m}^{\text{RN}}\right)\,Y_{l m}(\hat{n}),
\label{equ:cmb-nilc}
\end{eqnarray}
where the harmonic coefficients residual foreground ($a_{l m}^{\text{RFG}}$)
and residual noise($a_{l m}^{\text{RN}}$) are given by:
\begin{eqnarray}
a_{l m}^{\text{RFG}}=\sum_{j}\sum_{k}\sqrt{\lambda_{j
    k}}\,\beta_{j k}^{\text{RFG}}\,h_{l}^{j}\,Y_{l m}(\xi _{j k})
\end{eqnarray}
and
\begin{eqnarray}
a_{l m}^{\text{RN}}=\sum_{j}\sum_{k}\sqrt{\lambda_{j
    k}}\,\beta_{j k}^{\text{RN}}\,h_{l}^{j}\,Y_{l m}(\xi _{j k}).
\end{eqnarray}
Equation \ref{equ:cmb-nilc} implies that the NILC estimate of CMB
contains some residual foreground and noise contamination.


\section{WMAP 7 year needlet ILC map}
\label{sec:wmap-ilc}

The WMAP satellite has observed the sky in five frequency bands
denoted K, Ka, Q, V and W, centred on the frequencies of 23, 33, 41,
61 and 94 GHz respectively. After $7$ years of observation, the
released data includes maps obtained with ten difference assemblies,
for $7$ individual years. One map is available, per year, for each of
the K and Ka bands, two for the Q band, two for the V band and four
for the W band. These sky maps are sampled using the HEALPix
pixelisation scheme at a resolution level (nside~~1024), corresponding
to approximately $12$ million sky pixels.

In a first step, for each frequency band, we average all the
difference-assembly maps obtained at the same frequency. This yields
$5$ band-averaged maps. However, contrarily to the 7-year average maps
at nside~=~512 also provided by WMAP, these maps at nside~=~1024 are
not offset corrected. In order to determine the values of the offset
for a particular frequency band, we have used standard resolution
7-year band-average maps of WMAP as references. Offset values, for all
frequency bands, are obtained by determining the mean of the
difference between the band-averaged map being considered (degraded to
nside~=~512), and the released `reference' map. We implement the
needlet ILC on these offset-corrected maps.

In addition to these maps, we use three foreground templates in our
analysis: dust at 100 microns, as
obtained by \citet{1998ApJ...500..525S}, the 408 MHz synchrotron map
of \citet{1981A&A...100..209H}, and the composite all-sky H-alpha map
of \citet{2003ApJS..146..407F}. The offset-corrected WMAP maps and
fore-mentioned foreground templates are convolved/deconvolved, in
harmonic space, to a common beam resolution (corresponding to that of
the W frequency channel of WMAP 7-year release) before the
implementation of the needlet ILC on the map set.

\begin{figure}
  \includegraphics[scale=0.45,angle=0]{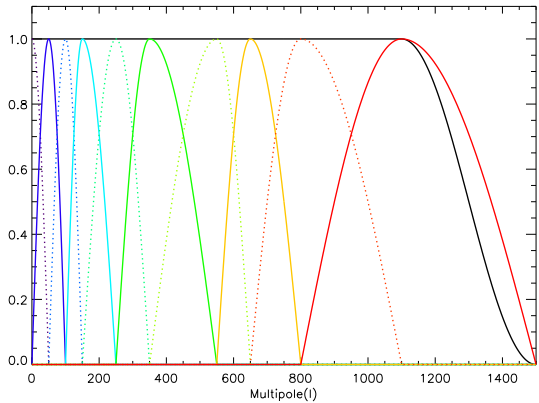}
  \caption{Needlet bands used in the present
    analysis. The black line shows the normalisation of the
    needlet bands, i.e. the total filter applied to the original map after
    needlet decomposition and synthesis of the output map from needlet coefficients.}
  \label{fig:needlet-bands}
\end{figure}

Each of these maps is decomposed into a set of needlet coefficients.
For each scale $j$, needlet coefficients of a given map are stored in
the format of a single HEALPix map at degraded resolution.  The
filters $h^{j}_{l}$ used to compute filtered maps are shaped as
follows:
\begin{eqnarray*} 
h^{j}_{l} = \left\{
\begin{array}{rl} 
\cos\left[\left(\frac{l^{j}_{peak}-l}{l^{j}_{peak}-l^{j}_{min}}\right)
\frac{\pi}{2}\right]& \text{for } l^{j}_{min} \le l < l^{j}_{peak},\\ 
\\
1\hspace{0.5in} & \text{for } l = l_{peak},\\
\\
\cos\left[\left(\frac{l-l^{j}_{peak}}{l^{j}_{max}-l^{j}_{peak}}\right)
\frac{\pi}{2}\right]& \text{for } l^{j}_{peak} < l \le l^{j}_{max} 
\end{array} \right. 
\end{eqnarray*}
For each scale $j$, the filter has compact support between the
multipoles $l^{j}_{min}$ and $l^{j}_{max}$ with a peak at
$l^{j}_{peak}$ (see figure \ref{fig:needlet-bands} and table
\ref{tab:needlet-bands}).  The needlet coefficients $\beta_{j k}$ are
computed from these filtered maps on HEALPix grid points $\xi_{j k}$
with resolution parameter nside equal to the smallest power of $2$
larger than $l^{j}_{max}/2$.

\begin{table}
\caption{List of needlet bands used in the present
  analysis.}  \centering \begin{tabular}{c c c c c} \hline\hline
  Band index & $l_{min}$ & $l_{peak}$ & $l_{max}$ & nside \\ [5ex]
  \hline 1 & 0 & 0 & 50 & 32\\ 2 & 0 & 50 & 100 & 64 \\ 3 & 50 & 100 &
  150 & 128 \\ 4 & 100 & 150 & 250 & 128 \\ 5 & 150 & 250 & 350 & 256
  \\ 6 & 250 & 350 & 550 & 512 \\ 7 & 350 & 550 & 650 & 512 \\ 8 & 550
  & 650 & 800 & 512 \\ 9 & 650 & 800 & 1100 & 1024 \\ 10 & 800 & 1100
  & 1500 & 1024 \\ [1ex] \hline
\end{tabular} 
\label{tab:needlet-bands} 
\end{table}

The estimates of needlet coefficients covariance matrices, for each
scale $j$, are computed by smoothing all possible products of needlet
coefficient maps $\beta^{c}_{j k}\beta^{c^{\prime}}_{j k}$ with
Gaussian beams. In this way, an estimate of needlet covariances at
each point $k$ is obtained as a local, weighted average of needlet
coefficient products. The process is summarised by the diagram:
\begin{eqnarray*}
\{\beta^{c}_{j k}\beta^{c^{\prime}}_{j k}\}_{k=1,..,N_{pix}}\,\xrightarrow{\text{SHT}}\,
\{a^{c\,c^{\prime}}_{lm}\}_{l=0,l^{j}_{max}, m=-l,..,0,..,l}\nonumber\\
\Big\downarrow{\times}\hspace{1.0in}\\
\{\hat{R}^{c\,c^{\prime}}_{j k}\}_{k=1,..,N_{pix}}\xleftarrow{\text{SHT}^{-1}}\,
\{B^{j}_{l}a^{c\,c^{\prime}}_{lm}\}_{l=0,l^{j}_{max}, m=-l,..,0,..,l}
\end{eqnarray*}
The full width at half maximum (FWHM) of each of the Gaussian windows 
used for this purpose is chosen to ensure the computation of
the statistics by averaging about 1200 samples or more. Choosing a
smaller FWHM results in excessive error in the covariance estimates,
and hence excessive bias. Choosing a larger FWHM results in less
localisation, and hence some loss of effectiveness of the needlet
approach.

\begin{figure}
  \centering
  \includegraphics[scale=0.3,angle=90]{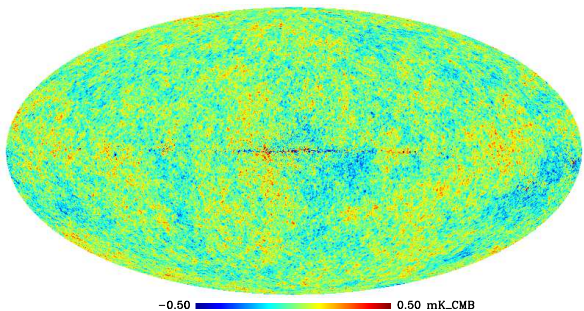}
  \caption{The NILC estimate of CMB temperature anisotropies 
     obtained by implementing the NILC on the 7-year band average
    maps (at nside~=~1024) provided by the WMAP collaboration.}
  \label{fig:cmb-map7yr}
\end{figure}

Using these covariance matrices, the ILC solution for each scale is
implemented locally to get the needlet weights and hence the estimated
CMB needlet coefficients. Finally, a full sky CMB map, displayed in
figure \ref{fig:cmb-map7yr}, is synthesised from these estimated
needlet coefficients. 

\section{CMB angular power spectrum}
\label{sec:wmap-spec}

Component separation with a needlet ILC is the first step in our
analysis. The cleaned CMB map, however, still contains residuals of
noise and foreground emission, and is impacted by the ILC bias due to
empirical correlations of CMB and foregrounds as well as by the effect
of calibration uncertainties and beam uncertainties.  Residual
contamination is clearly visible, for instance, along a narrow strip
on the Galactic plane of the map displayed in figure
\ref{fig:cmb-map7yr}.

In the following, we discuss these various types of errors, and our
strategy to minimise or evaluate their impact on the estimation of the
CMB power spectrum.

\subsection{Impact of instrumental noise}

It would be possible to use the estimated noise level in the WMAP maps
to de-bias the power spectrum of the map obtained in section
\ref{sec:wmap-ilc} from instrumental noise contribution. This,
however, requires, in particular at high multipoles, a very accurate
estimate of the noise level.

Thus, instead of trying to de-bias from instrumental noise a single power spectrum 
(computed on a single NILC map averaging all seven years of data),
we produce independent CMB maps for each of the
individual $7$ years of observations, and compute the CMB power
spectrum as an average of the $21$ cross power spectra obtained from the
$7$ clean CMB maps. In practice, all $7$ maps are obtained using
the same set of needlet weights, determined using the co-added 7 year
observations. Instrumental noise, being uncorrelated from year to
year, does not bias the average power spectrum, and can be ignored in
the estimate.

\subsection{Impact of residual foregrounds}

As the ILC weights used for all years are the same, residual
foreground emission will be the same in all maps. These residuals,
although small (as compared to original foreground contamination in
the maps of the individual differencing assemblies), are thus 100\%
correlated between the different single year maps, and bias the CMB
power spectrum. They are dealt with using a combination of masking and
de-basing, as follows.

First of all, the impact of most of the contamination from residuals
of bright point sources is removed from our estimate of power spectra
by applying to our output CMB maps, for each individual year, 
the point source mask provided by the WMAP collaboration,
and filling-in the holes by an interpolation procedure 
using the values of CMB anisotropies in the neighbouring
unmasked pixels. We start with the masked border pixels, assign to
each of them the average of all observed pixels which are within a
distance of two pixel sizes. Then we iterate, increasing the distance
for averaged pixels by two pixel sizes at each iteration. While not
theoretically optimal, this procedure works very well in practice.

The impact of galactic contamination on our power spectrum estimate is
limited by the use of a conservative galactic mask,
$M_{\bar{\theta}_{r},\bar{\theta}_{w}}$(applied after the NILC, so that we
still get full sky maps). Our mask simply excludes sky regions in
which galactic residuals after component separation can be too strong for
precise CMB power spectrum estimation. The apodised mask used in
practice is: 
\begin{eqnarray*} 
M_{\bar{\theta}_{r},\bar{\theta}_{w}}(\theta,\phi)  = \left\{
\begin{array}{rl} 
0\hspace{0.5in}& 
\text{for } 0^{\circ} \le \left|\bar{\theta}\right| < \bar{\theta}_{r},
\\ \\\sin^{2}\left[\left(\frac{\bar{\theta}-\bar{\theta}_{r}}{\bar{\theta}_{w}}\right)
\frac{\pi}{2}\right]& 
\text{for  } \bar{\theta}_{r} \le \left|\bar{\theta}\right| \le \bar{\theta}_{r}+\bar{\theta}_{w},
\\ \\ 1\hspace{0.5in}
& \text{for } \left|\bar{\theta}\right| > \bar{\theta}_{r}+\bar{\theta}_{w}\\
\end{array} \right. 
\label{equ:apod-mask}
\end{eqnarray*}
\begin{figure}
  \includegraphics[scale=0.45,angle=0]{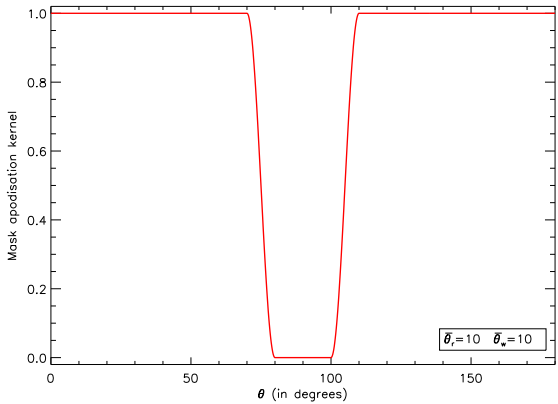}
  \caption{Apodisation kernel for Galactic mask used in our analysis}
  \label{fig:apod-mask}
\end{figure}

\noindent
where
$\bar{\theta}=\left(\frac{\pi}{2}-\theta\right)\frac{180^{\circ}}{\pi}$
is the latitude. The apodisation of the mask borders limits the aliasing of large scales into small scales.

\subsection{Validation on simulations and de-biasing}
The method is tested, tuned, and validated using numerical simulations
based on the Planck Sky Model (PSM). A general description of the PSM
can be found on the PSM web page,\footnote{\small
  http://www.apc.univ-paris7.fr/$\sim$delabrou/psm.html} as well as in
\citet{2008A&A...491..597L} and/or in \citet{2009A&A...503..691B}. The
simulated sky emission comprises:
\begin{itemize}
\item a Gaussian CMB temperature map, with a power spectrum matching
the WMAP best fit;
\item galactic emission, consisting in the superposition of
synchrotron, free-free, thermal dust, and spinning dust;
\item SZ effects, both thermal and kinetic; 
\item a population of point sources.
\end{itemize}

\noindent
Noise maps for the individual years are randomly generated with a
Gaussian probability distribution that is uncorrelated from pixel to
pixel and between different years.  The noise variance per pixel is
inversely proportional to the hit count for each individual year.
This is expected to describe reasonably accurately the expected noise
behaviour of the actual WMAP data. The noise level impacts the exact
value of the NILC coefficients (in particular at high galactic
latitude and on small scales), and the exact size of the error bars in
the measured CMB power spectrum. Small uncertainties in the noise,
however, do not bias the estimated $C_\ell$.

Biases are investigated in the following way.  Fifty different
realisations of the full simulations are generated, and processed in
the exact same way as the actual WMAP data sets. The average
systematic difference (i.e. bias) between the recovered power spectra
and the input CMB power spectrum is computed. The bias estimated in
this way (displayed with green right-facing triangles in figure
\ref{fig:cmb-cl-bias}) is typically less than $3-5$\% of the estimated
$C_{l}$, and also typically smaller than the total errors on
individual $C_{l}$ bins (figure \ref{fig:cmb-cl-bias}). This average
bias is nonetheless corrected for in our estimate of the CMB power
spectrum (see section \ref{sec:est-pow-spec} for details).

The observed bias is due to the combination of two effects:
additive residual foreground emission, which generates a positive
bias, and empirical correlations between contaminants (foregrounds
and noise) and CMB maps which generate the negative ILC bias discussed
at length in \citet{2009A&A...493..835D}.

Our de-biasing procedure deserves a discussion, since
using this average bias estimated on simulations to de-bias (by
subtraction) the actual CMB power spectrum obtained on real WMAP data
is valid only if the simulations are representative of the real sky,
i.e.

\begin{itemize}
\item the CMB power spectrum assumed in the simulations must be close
  to the real one;
\item the simulated foregrounds must be representative of the sky
  (similar levels and behaviour)
\item the description of the instrument must be correct.
\end{itemize}
The real CMB power spectrum and the real foregrounds are not known to
infinite accuracy (otherwise there would be no point in trying to
measure them again). However, the following points give us confidence
in the representativeness of our simulations:
\begin{itemize}
\item Concerning the CMB, a small error $\Delta
C_l$ in the simulated CMB $C_l$ would generate a miscalculation of the
multiplicative ILC bias. The error in this bias estimate, however, is
of second order (of order 2\% times $\Delta C_l$), with a minor impact
on the $C_l$ estimate. 
\item The galactic foreground model in the PSM is built on the basis
  of the WMAP foreground analysis of \citet{2008A&A...490.1093M}. Even
  if the model has known limitations and uncertainties, it yields sky
  emission in the WMAP channels that is in good agreement with the
  observations.  For safety in our power spectrum estimation however,
  a conservative mask being applied to exclude the galactic plane
  region from the calculation of temperature $C_l$.
\item PSM point sources are based on real observations
  of radio sources with GB6 \citep{1998astro.ph.12237G}, PMN
  \citep{1994ApJS...90..179G,1994ApJS...91..111W,
    1995ApJS...97..347G,1996ApJS..103..145W}, SUMSS
  \citep{2003MNRAS.342.1117M} and NVSS \citep{1998AJ....115.1693C},
  and includes the full catalog of WMAP detected point sources. 
  The PSM point source model is
  hence thought to be reasonably representative of the expected point
  source contamination in WMAP data sets.
\item The thermal and kinetic SZ effects are very faint as compared to
  WMAP sensitivity, and do not play a significant role in this
  analysis.
\item Uncertainties connected to the description of the instrument
  that are susceptible to impact our simulations are mostly
  calibration uncertainties and beam uncertainties. The impact of
  these, which turns out to be negligible as estimated on simulations,
  will be discussed later on.
\end{itemize}

\begin{figure}
  \includegraphics[scale=0.45,angle=0]{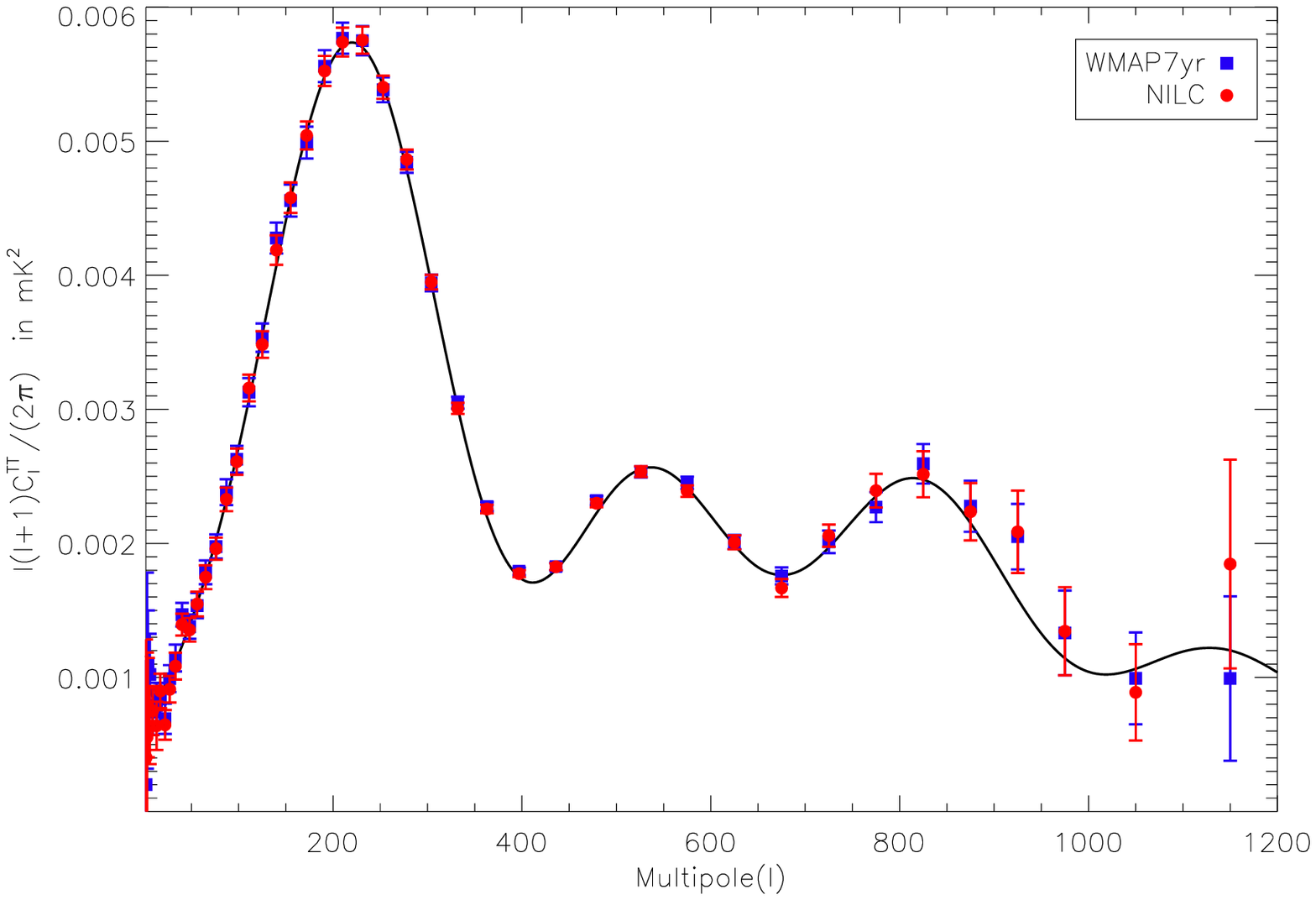}
  \includegraphics[scale=0.45,angle=0]{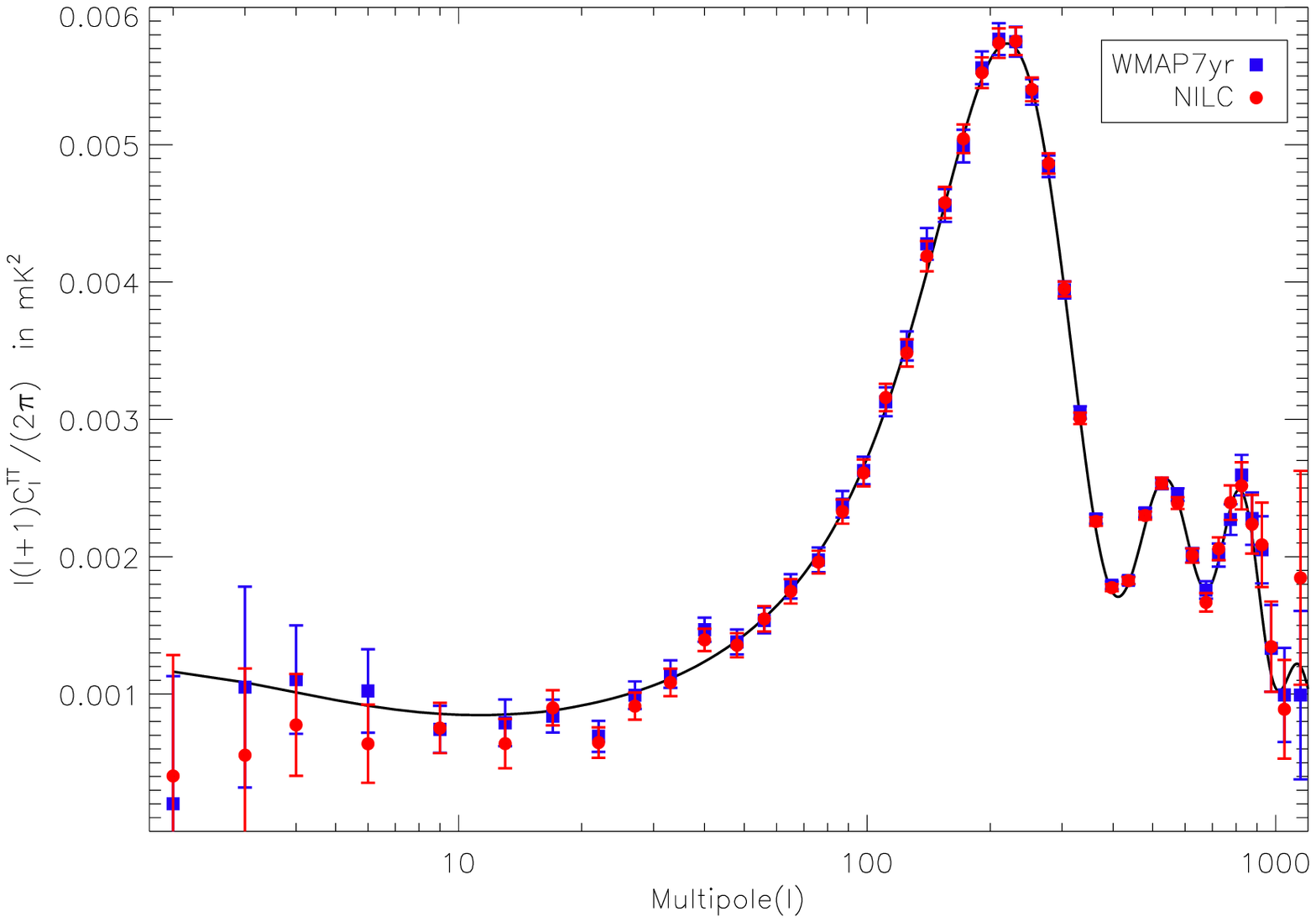}
  \caption{The red filled circles show the angular power spectrum
    estimated, after foreground subtraction with the NILC, using $7$
    years of observations of WMAP. The blue filled squares show the
    7-year angular power spectrum published by the WMAP
    collaboration. The black solid line shows the theoretical angular
    power spectrum for WMAP best-fit Lambda-CDM model. The top panel
    uses a linear scale in the horizontal axis, and the bottom panel a
    logarithmic scale.  Our estimate is consistent with that of WMAP,
    although it has significantly less power at low multipoles
    ($l<15$).}
  \label{fig:cmb-cl}
\end{figure}
\begin{figure}
  \includegraphics[scale=0.45,angle=0]{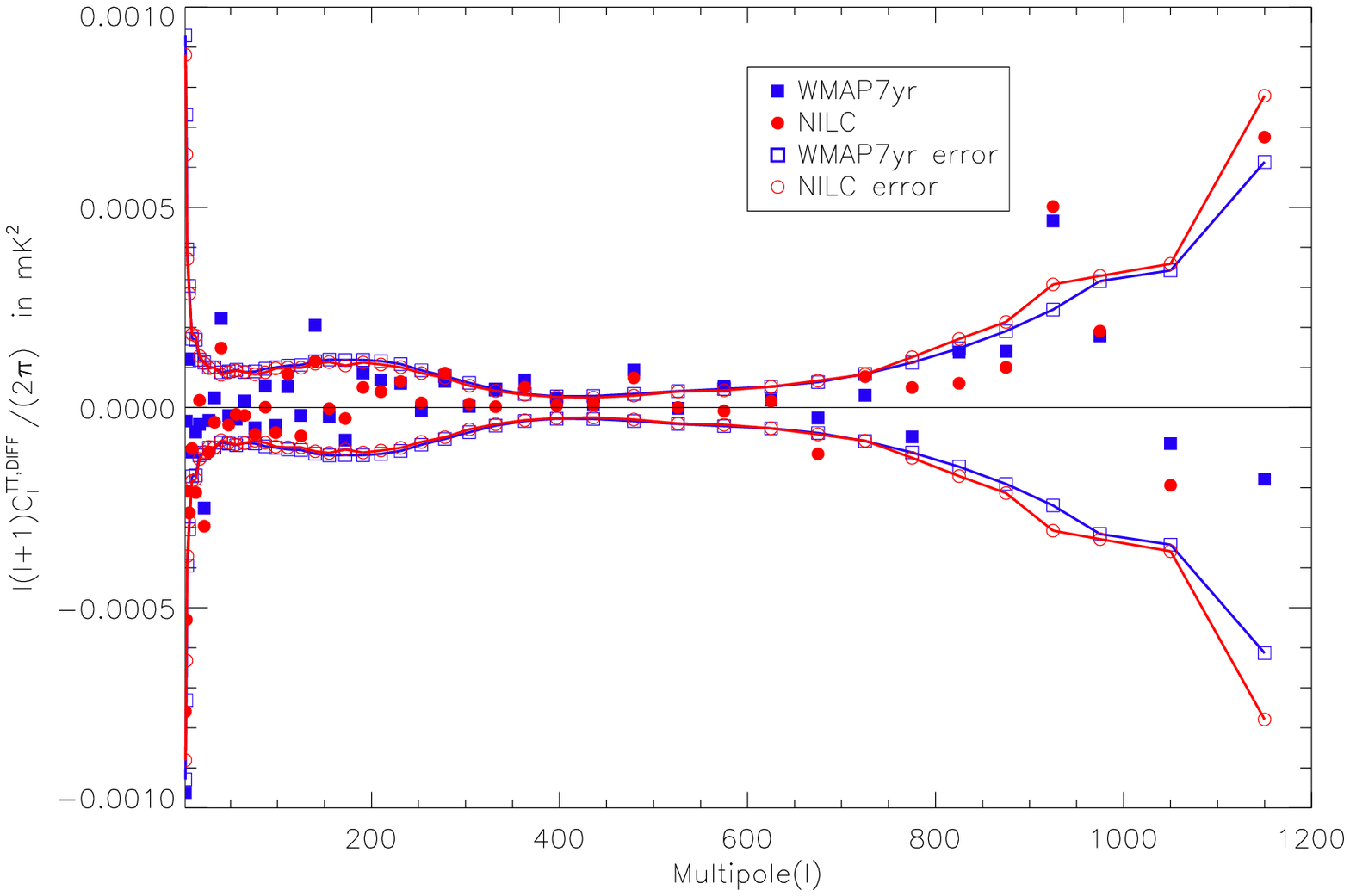}
  \includegraphics[scale=0.45,angle=0]{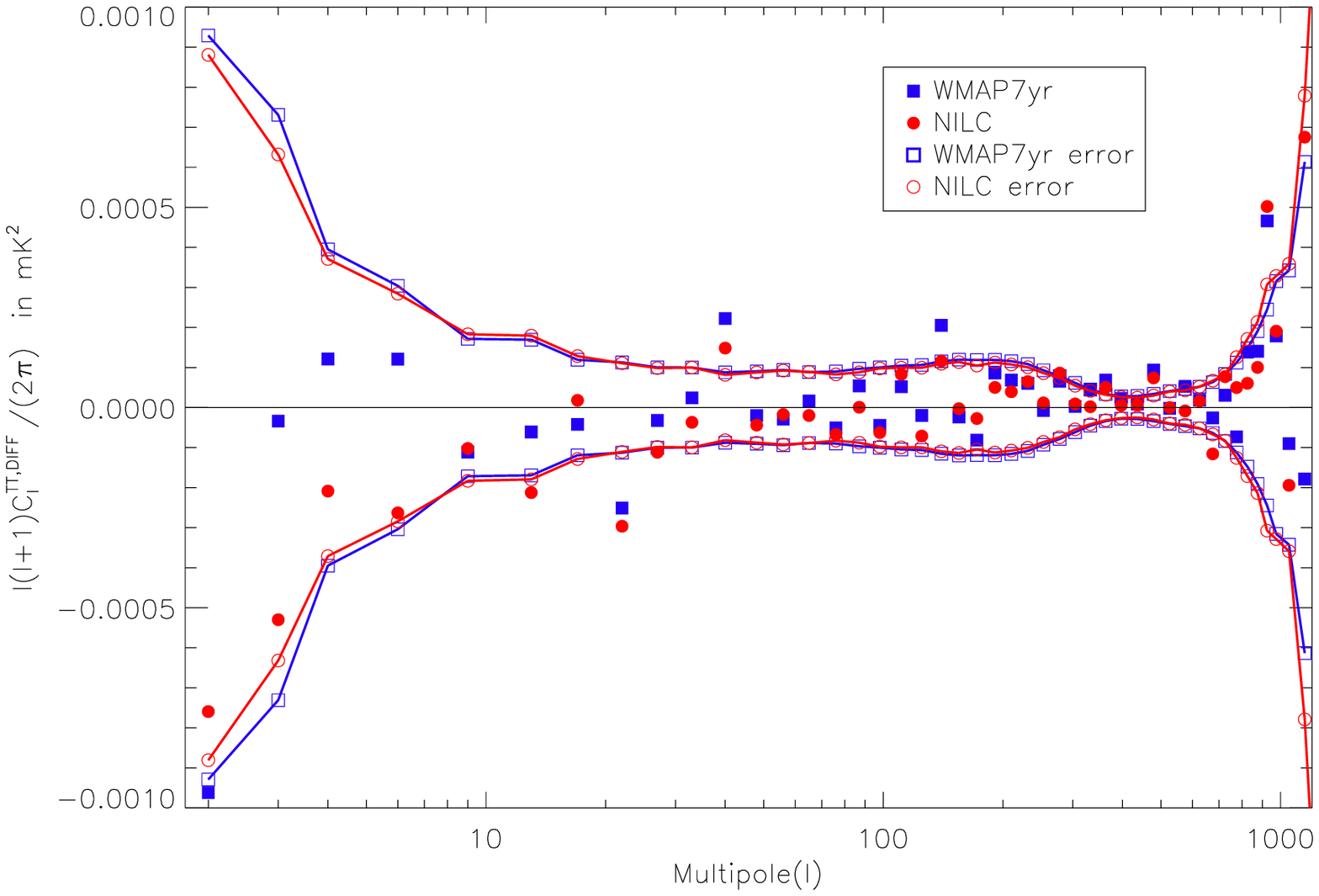}
  \caption{The red filled circles show the difference between the
    angular power spectrum estimated using NILC maps and the
    theoretical WMAP best fit angular power spectrum. The red open
    circles show the total error in the estimate of the angular power
    spectrum using our method. Similarly, the blue filled squares show
    the difference between the angular power spectrum published by the
    WMAP collaboration and theoretical best fit. The blue open squares
    show the error in the estimate of angular power spectrum provided
    by WMAP. The top panel uses a linear scale in the horizontal axis,
    and the bottom panel a logarithmic scale. }
  \label{fig:cmb-cl-diff}
\end{figure}

\subsection{Power spectrum estimation}
\label{sec:est-pow-spec}
Residual noise in our CMB maps is inhomogeneous, primarily because 
of non-uniform sky coverage, with higher number of
observations in the directions of ecliptic poles and rings around them
of $45^{\circ}$ radius. Taking into account this inhomogeneity for computing 
the CMB power spectrum is expected to yield more precise estimates of
the CMB $C_l$ in multipoles where the noise is the main source of error.

We follow the general idea of the noise-weighting using needlets
described in \citet{2008PhRvD..78h3013F}, with a number of changes to
adapt the method to our present analysis.  Consider a bin in $l$,
defined by a window $H_b(l)=\left [h_b(l) \right] ^2$ restricted to
the range in $l$ limited by $l_{\rm min}^b$ and $l_{\rm max}^b$, such
that
\begin{eqnarray*} 
H_{b}(l)  = \left\{
\begin{array}{cl} 
\frac{l(l+1)}{2\pi}\frac{1}{(2l+1)}\frac{1}{(\Delta l)_{b}} & \;\;\; l_{min}^b\le l \le l_{max}^{b}
\\ \\ 0 & \;\;\;  \text{otherwise}\\
\end{array} \right. 
\end{eqnarray*} 
with $(\Delta l)_b = l_{\rm max}^b-l_{\rm min}^b+1$.
The power $P_b$ of the CMB map filtered with $h_b(l)$ is an estimate of the CMB
angular power spectrum $C_\ell$ in the 
window $H_b(l)$, with:
\begin{eqnarray}
P_b & = & \sum_l H_b(l) (2l+1) C_l \nonumber \\
& = & \frac{1}{(\Delta l)_{b}} \sum_l \frac{l(l+1)}{2\pi} C_l
\label{eq:Pb}
\end{eqnarray}
Now starting from a pair of reconstructed CMB maps for two different
years of observation $i$ and $j$, filtering each of them by $h_b(l)$,
we can form quadratic estimators of the band-averaged power spectrum
between $l_{\rm min}$ and $l_{\rm max}$:
\begin{eqnarray}
\widehat P_b^{ij}=\frac{4\pi}{N_{\rm pix}}\sum_{\xi}T_{b,i}(\xi)T_{b,j}(\xi)
\end{eqnarray}
where $T_{b,i}(\xi)$ is the NILC estimate of CMB map for year $i$,
filtered in the band $h_b(l)$, and $N_{\rm pix}$ is the number of
pixels in the filtered map.

In fact, for each single pixel $\xi$, the quantity 
\begin{eqnarray}
\widehat P_b^{ij}(\xi)=4\pi\,T_{b,i}(\xi) T_{b,j}(\xi)
\end{eqnarray}
is itself an estimator of $P_b$,. We form a noise-weighted average of
the estimators for all pixels and all pairs of maps of the form:
\begin{eqnarray}
\widehat P_b=\frac{1}{N_{\rm pix}}\frac{2}{N_{\rm year}\,(N_{\rm year}-1)}\sum_\xi \sum_{i,j,\,i<j} W_b^{ij}(\xi)\widehat P_b^{ij}(\xi)
\end{eqnarray}
using weights proportional to
\begin{eqnarray}
W_{b}^{ij}(\xi) \propto 1\Bigg{/}\left[\left(\frac{1}{4\pi}\,P_{b}+\sigma^2_{\rm noise}(\xi)\right)^{2}+\frac{\sigma^4_{\rm noise}(\xi)}{N_{\rm year}-1}\right]
\end{eqnarray}
where $N_{\rm year}$ is the number of years of observation and
$\sigma^2_{\rm noise}(\xi)$ is the variance of the pixel noise of the
reconstructed 7-year NILC CMB map filtered by $h_b(l)$. The weights are
normalised so that no power is lost:
\begin{eqnarray}
\frac{1}{N_{\rm pix}}\frac{2}{N_{\rm year}\,(N_{\rm year}-1)}\sum_\xi \sum_{i,j,\, i<j} W_b^{ij}(\xi) & = & 1.
\end{eqnarray}
These weights are very similar to those in equation 18 of
\citet{2008PhRvD..78h3013F}, used for for power spectrum estimation
using a needlet frame, except that here we have a correction term
${\sigma^4_{\rm noise}(\xi)}/{(N_{\rm year}-1)}$, coming from the fact
that only cross spectra of the $N_{\rm year}$ maps are used in our
estimator.  When a large number of independent maps are used, this
term is small and can be neglected (in our case, with $N_{\rm year}
=7$, the correction to the weights is of order 15\% only).

The value of $\sigma^2_{\rm noise}(\xi)$ is estimated using 100
independent realisations of the reconstruction of the CMB by the NILC,
and $P_{b}$ is estimated by plugging in the WMAP best fit $C_l$ in
equation \ref{eq:Pb}. Note that some imprecision on $P_b$ and
$\sigma^2_{\rm noise}$ has little impact on the estimation. In case of
errors in the estimation of $P_b$ and $\sigma^2_{\rm noise}$, the
weights used are not perfectly optimal, but this induces no bias in
the estimator.

An apodised galactic mask $M_{10^{\circ},10^{\circ}}$ is used
in the power spectrum estimation (see equation (\ref{equ:apod-mask})
and figure \ref{fig:apod-mask}). Such masking is equivalent to
lowering the weight of some of the pixels (down to vanishing
weights at the lowest galactic latitudes).

\begin{figure}
  \includegraphics[scale=0.45,angle=0]{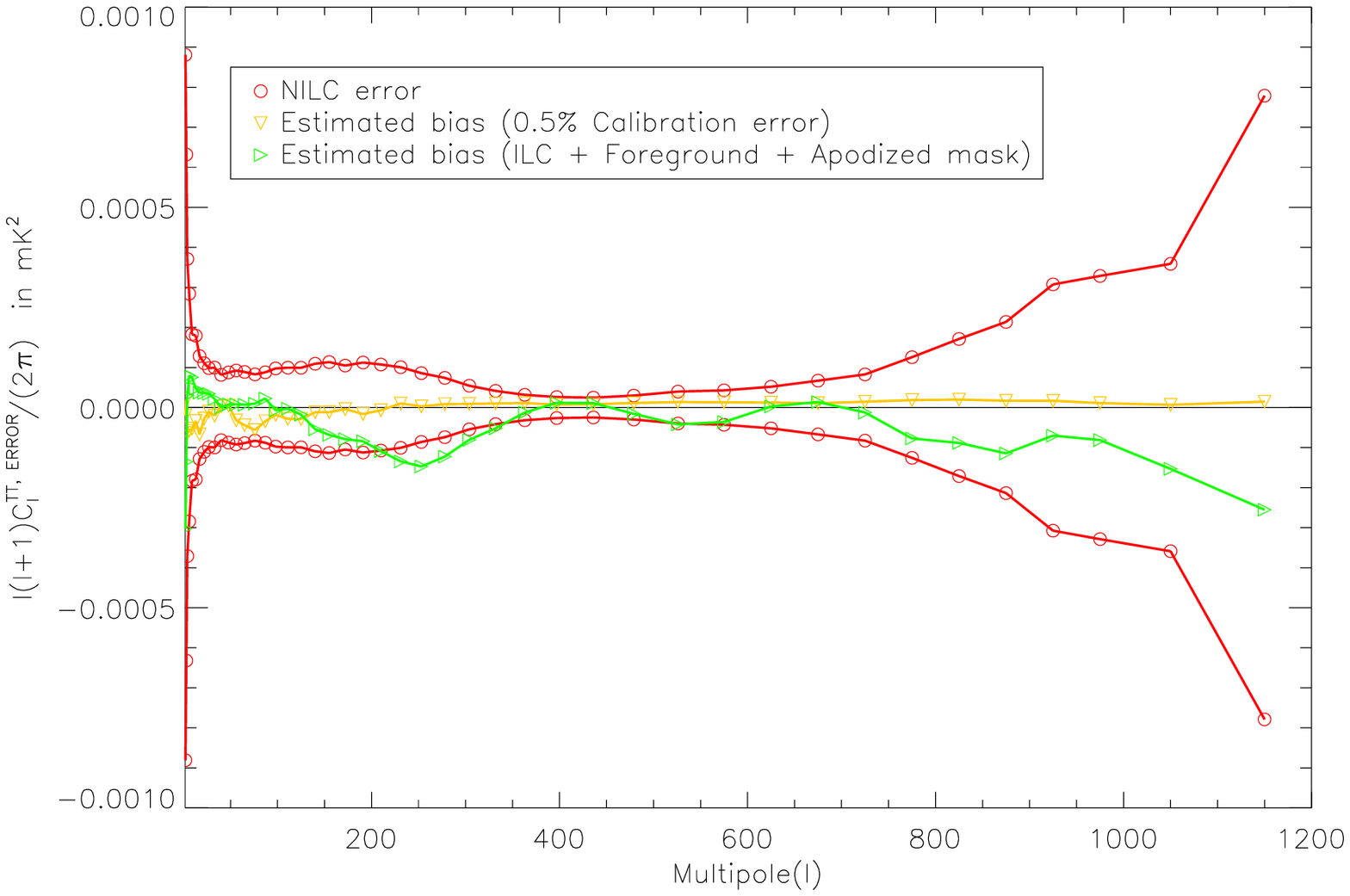}
   \includegraphics[scale=0.45,angle=0]{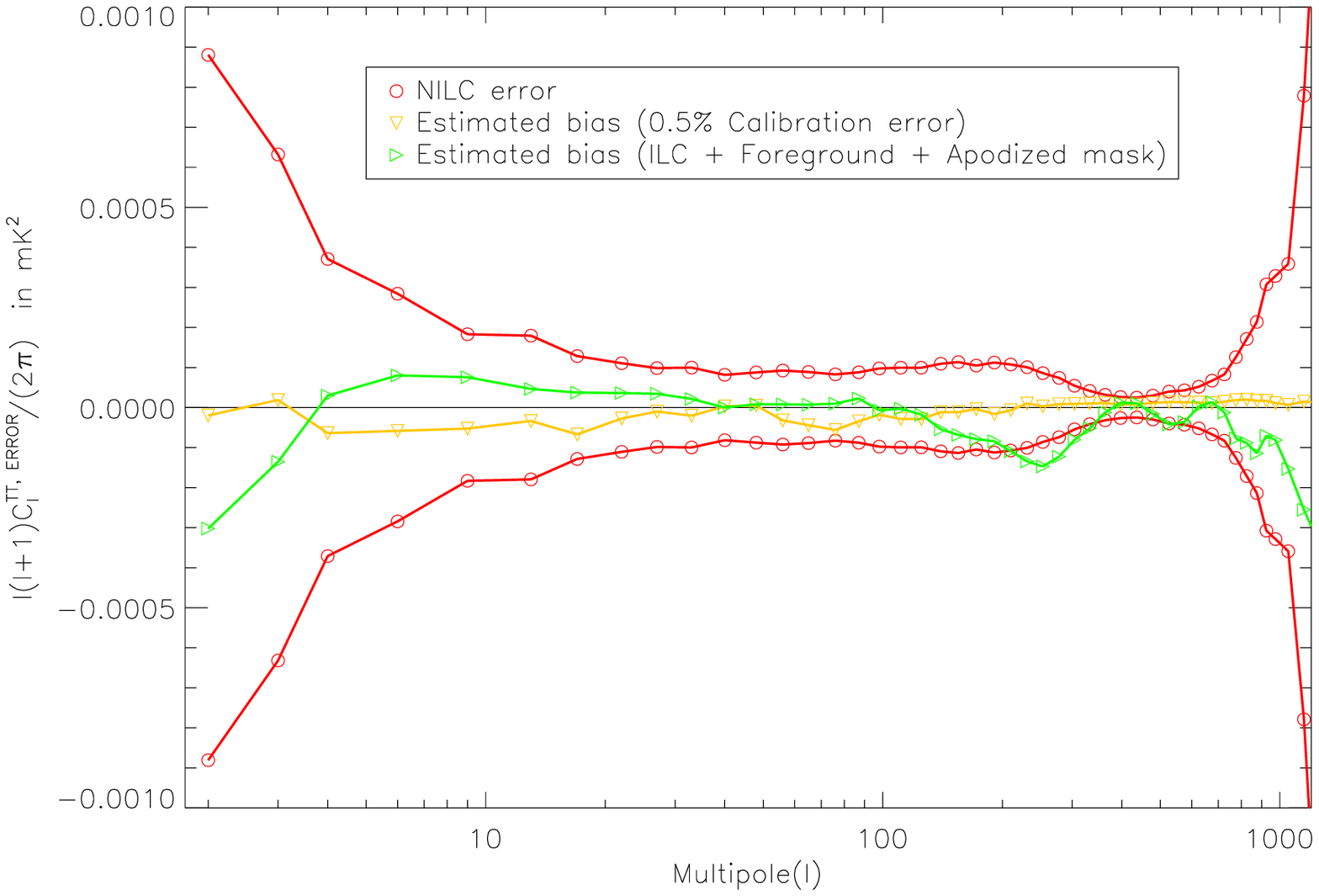}
 \caption{The green right-facing triangles show the bias estimated by
   implementing the NILC on 50 WMAP-like simulated data. The yellow
   downward triangles show the estimate of the bias due to calibration
   error of the order of of $0.5\%$. The red open circles show the
   total error in the estimate of angular power spectrum using
   NILC. The top panel uses a linear scale in the horizontal axis, and
   the bottom panel a logarithmic one.  }
  \label{fig:cmb-cl-bias}
\end{figure}

Figure \ref{fig:cmb-cl} shows the binned NILC estimate of the angular
power spectrum of CMB temperature anisotropies, after subtracting the
binned ILC bias. The estimate of the same obtained by the WMAP
collaboration is plotted on the same figure for comparison.  Figure
\ref{fig:cmb-cl-diff} shows, for each of them, the difference with the
theoretical angular power spectrum, and figure \ref{fig:cmb-cl-bias}
displays the estimated NILC and calibration biases, compared to the
statistical error of the estimator.

The error bars in the estimated power spectrum include the statistical
error of the estimator (noise and cosmic variance terms) and the
statistical error on the ILC bias term -- the value of which is known
only up to statistical error, by reason of the intrinsic variance of
the empirical correlation between the CMB and the contaminants
(foregrounds and noise).  The power spectrum obtained using our
analysis agrees well with that provided by the WMAP collaboration and
is consistent with the $\Lambda$CDM model for WMAP best-fit
cosmological parameters, except for a small but significant lack of
power at low multipoles, as can be seen in figure \ref{fig:cmb-cl}. We
do not observe such systematic differences on our analysis of
simulated data sets.

\begin{figure}
  \includegraphics[scale=0.45,angle=0]{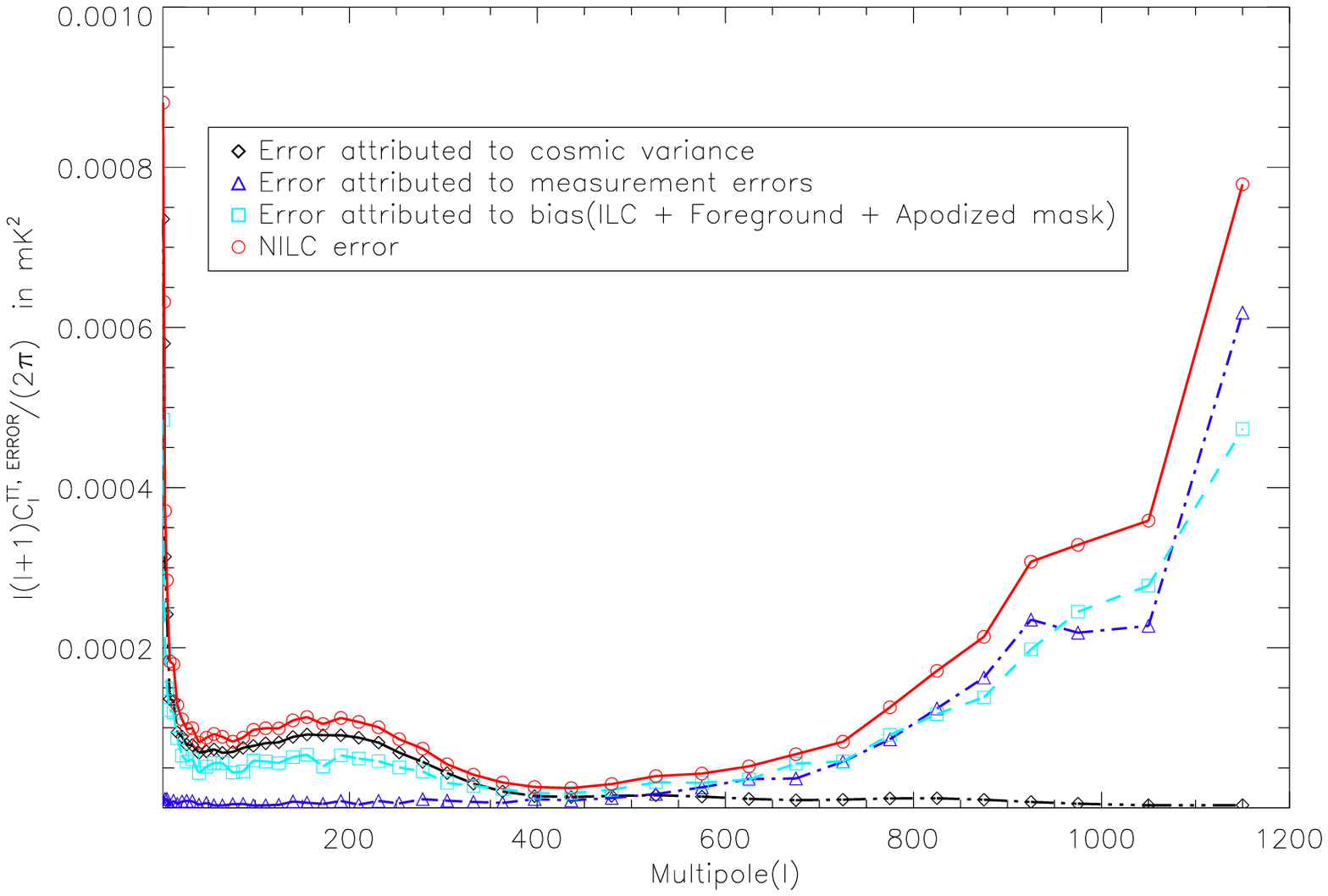}
  \includegraphics[scale=0.45,angle=0]{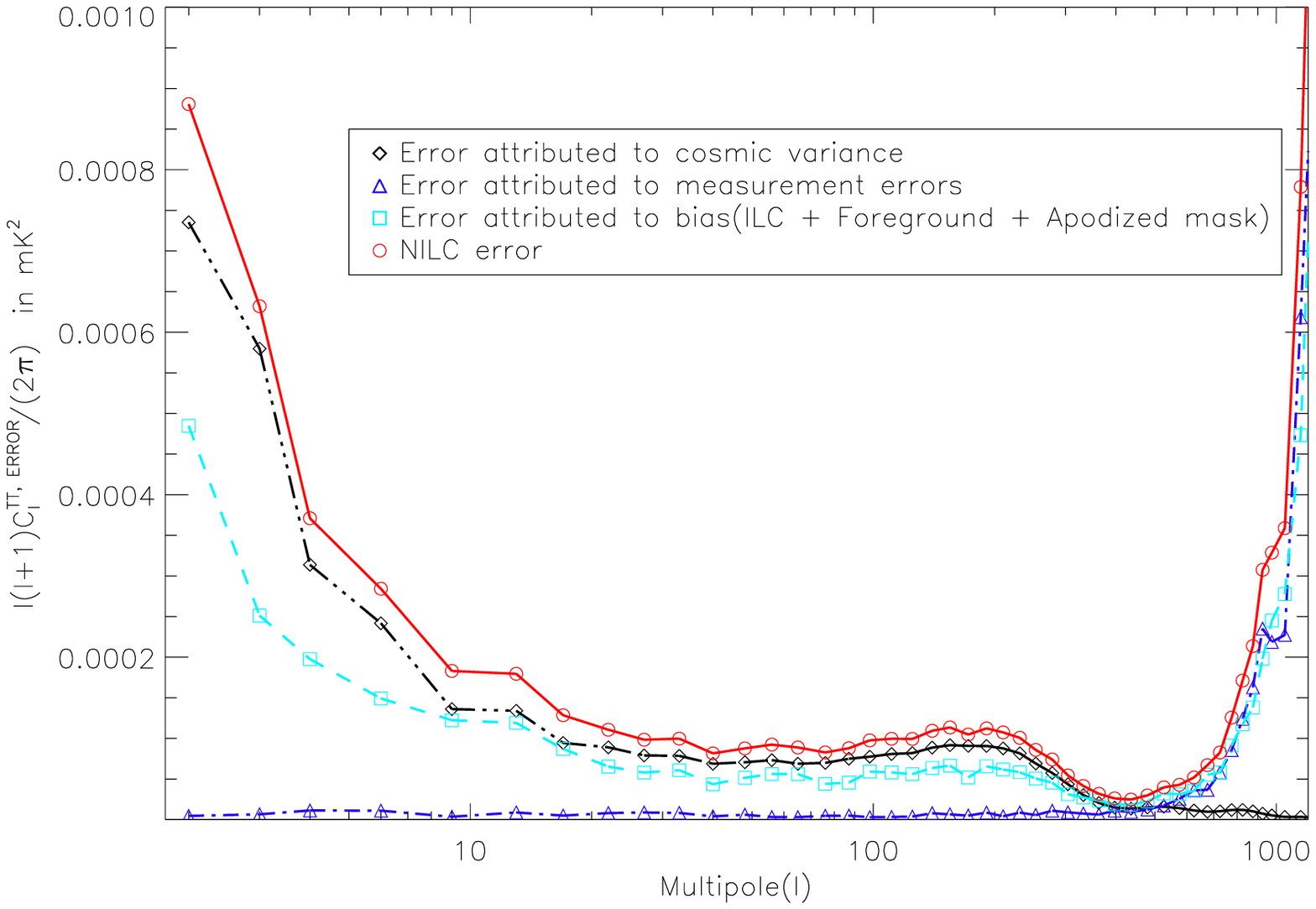}
  \caption{The red circles show the total error in the estimate
    of the  CMB angular power spectrum using the NILC. The black open diamonds show the
    error due to cosmic variance. The blue upward triangles show the
    measurement error. The sky blue open squares show the error due
    to bias uncertainty. The horizontal axis of the
    top plot is scaled linearly, and that of the bottom plot is scaled
    logarithmically.}
  \label{fig:cmb-cl-err}
\end{figure}
\begin{table}
\caption{Comparison of NILC estimate of binned angular power spectrum
  (after bias correction) with the estimate of binned
  angular power spectrum provided by WMAP collaboration.}
\centering \begin{tabular}{c c c c c} \hline\hline $l_{b}$ &
  $P_{b,\,\text{NILC}}$ & $P_{b,\,\text{WMAP}}$ &
  $P^{\text{ERR}}_{b,\,\text{NILC}}$ &
  $P^{\text{ERR}}_{b,\,\text{WMAP}}$\\ $ $ & $ $ & $ $ & $ $ & $ $\\ $
  $ & $(mK^{2})$ & $(mK^{2})$ & $(mK^{2})$ & $(mK^{2})$ \\ [1ex]
  \hline 2 & 4.033e-04 & 2.009e-04 & 8.807e-04 & 9.291e-04 \\ 3 &
  5.547e-04 & 1.051e-03 & 6.319e-04 & 7.308e-04 \\ 4 & 7.754e-04 &
  1.105e-03 & 3.710e-04 & 3.949e-04 \\ 6 & 6.380e-04 & 1.022e-03 &
  2.843e-04 & 3.039e-04 \\ 9 & 7.523e-04 & 7.432e-04 & 1.831e-04 &
  1.715e-04 \\ 13 & 6.391e-04 & 7.906e-04 & 1.795e-04 & 1.693e-04
  \\ 17 & 8.995e-04 & 8.396e-04 & 1.285e-04 & 1.193e-04 \\ 22 &
  6.469e-04 & 6.922e-04 & 1.107e-04 & 1.128e-04 \\ 27 & 9.119e-04 &
  9.915e-04 & 9.843e-05 & 1.004e-04 \\ 33 & 1.085e-03 & 1.146e-03 &
  9.975e-05 & 9.997e-05 \\ 40 & 1.394e-03 & 1.468e-03 & 8.158e-05 &
  8.780e-05 \\ 48 & 1.355e-03 & 1.379e-03 & 8.772e-05 & 9.041e-05
  \\ 56 & 1.548e-03 & 1.536e-03 & 9.239e-05 & 9.395e-05 \\ 65 &
  1.748e-03 & 1.784e-03 & 8.884e-05 & 8.840e-05 \\ 76 & 1.961e-03 &
  1.977e-03 & 8.277e-05 & 8.999e-05 \\ 87 & 2.329e-03 & 2.382e-03 &
  8.794e-05 & 9.679e-05 \\ 98 & 2.610e-03 & 2.628e-03 & 9.761e-05 &
  1.002e-04 \\ 111 & 3.159e-03 & 3.128e-03 & 9.954e-05 & 1.046e-04
  \\ 125 & 3.483e-03 & 3.535e-03 & 9.936e-05 & 1.063e-04 \\ 140 &
  4.188e-03 & 4.278e-03 & 1.094e-04 & 1.153e-04 \\ 155 & 4.578e-03 &
  4.558e-03 & 1.135e-04 & 1.195e-04 \\ 172 & 5.043e-03 & 4.989e-03 &
  1.048e-04 & 1.187e-04 \\ 191 & 5.524e-03 & 5.560e-03 & 1.124e-04 &
  1.190e-04 \\ 210 & 5.739e-03 & 5.768e-03 & 1.074e-04 & 1.157e-04
  \\ 231 & 5.754e-03 & 5.750e-03 & 1.007e-04 & 1.085e-04 \\ 253 &
  5.402e-03 & 5.383e-03 & 8.588e-05 & 9.294e-05 \\ 278 & 4.864e-03 &
  4.843e-03 & 7.407e-05 & 7.856e-05 \\ 304 & 3.949e-03 & 3.943e-03 &
  5.446e-05 & 6.153e-05 \\ 332 & 3.007e-03 & 3.050e-03 & 4.148e-05 &
  4.507e-05 \\ 363 & 2.256e-03 & 2.275e-03 & 3.179e-05 & 3.374e-05
  \\ 397 & 1.775e-03 & 1.792e-03 & 2.620e-05 & 2.785e-05 \\ 436 &
  1.826e-03 & 1.828e-03 & 2.476e-05 & 2.863e-05 \\ 479 & 2.300e-03 &
  2.319e-03 & 2.983e-05 & 3.404e-05 \\ 526 & 2.535e-03 & 2.533e-03 &
  3.979e-05 & 4.099e-05 \\ 575 & 2.390e-03 & 2.451e-03 & 4.292e-05 &
  4.647e-05 \\ 625 & 2.008e-03 & 2.011e-03 & 5.195e-05 & 5.202e-05
  \\ 675 & 1.667e-03 & 1.757e-03 & 6.714e-05 & 6.344e-05 \\ 725 &
  2.058e-03 & 2.012e-03 & 8.267e-05 & 8.385e-05 \\ 775 & 2.394e-03 &
  2.271e-03 & 1.257e-04 & 1.123e-04 \\ 825 & 2.516e-03 & 2.594e-03 &
  1.712e-04 & 1.477e-04 \\ 875 & 2.237e-03 & 2.277e-03 & 2.137e-04 &
  1.908e-04 \\ 925 & 2.087e-03 & 2.051e-03 & 3.075e-04 & 2.446e-04
  \\ 975 & 1.345e-03 & 1.333e-03 & 3.286e-04 & 3.158e-04 \\ 1050 &
  8.890e-04 & 9.932e-04 & 3.588e-04 & 3.425e-04 \\ 1150 & 1.846e-03 &
  9.924e-04 & 7.788e-04 & 6.132e-04 \\ [1ex] \hline
\end{tabular} 
\label{tab:binned_spec} 
\end{table}

\subsection{Impact of calibration errors}
Calibration errors are a serious issue for precise separation of the
CMB from foregrounds using any type of ILC. As discussed by 
\citet{2010MNRAS.401.1602D}, they
`conspire' with the ILC filter to cancel out the CMB.
This effect is particularly strong  in the high signal
to noise ratio regime, which is the case in our present analysis, in
particular on large scales. We investigate the impact of this
calibration bias by redoing the analysis using slightly modified
calibration coefficients (differences of the order of $0.5\%$), and
computing the difference between the CMB spectra estimated in both
cases. The result, plotted in figure \ref{fig:cmb-cl-bias}, shows that
here even at large angular scale, the bias due to calibration error is small. 
We hence neglect this effect
in our estimate of errors on $C_\ell$.  Note however that this
calibration bias will be a concern for those experiments which measure
CMB with higher signal to noise ratio than WMAP. 

\subsection{Impact of beam uncertainties}
The uncertainty on beam shapes is equivalent to a calibration error
which depends on the harmonic mode $l$.  As this uncertainty is
essentially at high $l$, where the signal to noise ratio is worse than
on large scales, the bias, for small beam shape errors, is not expected to 
impact much the CMB
reconstruction here, and is also neglected in the present analysis.


\section{Discussion}
\label{sec:wmap-discuss}

\subsection{Low power on large scales?}

The temperature power spectrum we obtain seems to be, on large scales,
systematically lower than the theoretical `best fit' power spectrum
(for $\ell < 15$) and the WMAP measurement (for $\ell < 40$, except
for the quadrupole, see figure \ref{fig:cmb-cl}, bottom left panel,
and table \ref{tab:binned_spec_diff}).

Considering that error bars on these scales are cosmic variance
dominated, the difference between the power spectrum measured by WMAP
and by us \emph{from the same original data set} is problematic, and
deserves a discussion and additional investigation. A systematic shift
in the estimated power spectrum at all $l$ below 40 is bound to impact
the interpretation of the observations, and the inferred values and
limits for cosmological parameters. In the following, we investigate
various possible origins for the discrepancy.

\begin{table}
\caption{Difference between the NILC estimate of binned angular power
  spectrum and the angular power spectrum provided by the WMAP
  collaboration. Except for the quadrupole, for which our estimate is
  larger (after debiasing) and except the bins centred at $l=9$ and
  $l=27$, our estimated power spectrum is lower than the WMAP one.}
\centering \begin{tabular}{c c } \hline\hline $l_{b}$ &
  $P_{b,\,\text{NILC}} - P_{b,\,\text{WMAP}}$ $(mK^{2})$ \\ [1ex]
  \hline 2 & 2.024e-04 \\ 3 & -4.963e-04 \\ 4 & -3.296e-04 \\ 6 &
  -3.840e-04 \\ 9 & 0.091e-04 \\ 13 & -1.515e-04 \\ 17 & 0.599e-04
  \\ 22 & -0.453e-04 \\ 27 & -0.796e-04 \\ 33 & -0.610e-04 \\ 40 &
  -0.740e-04 \\ [1ex] \hline
\end{tabular} 
\label{tab:binned_spec_diff} 
\end{table}

\begin{figure}
  \includegraphics[scale=0.45,angle=0]{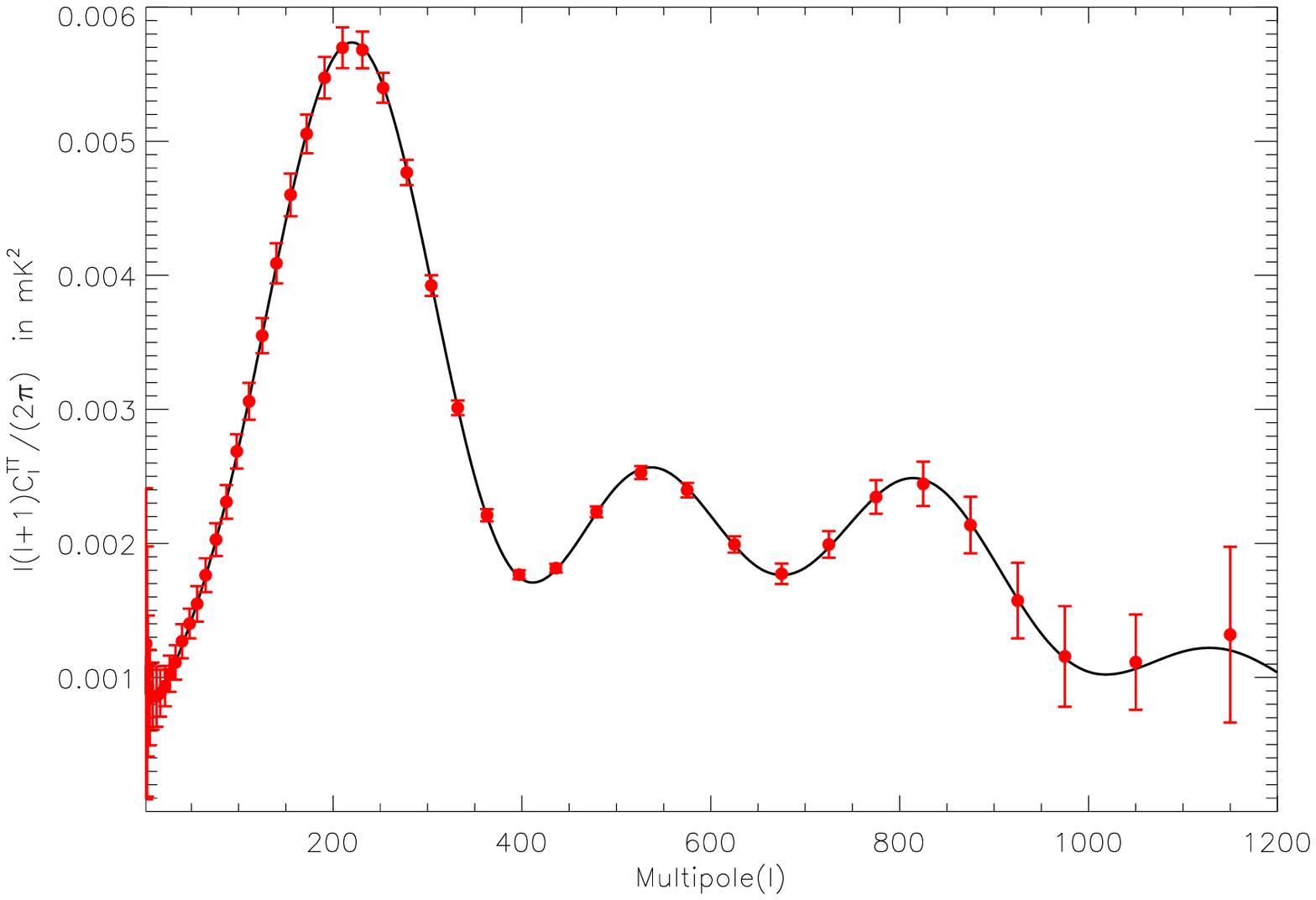}
  \includegraphics[scale=0.45,angle=0]{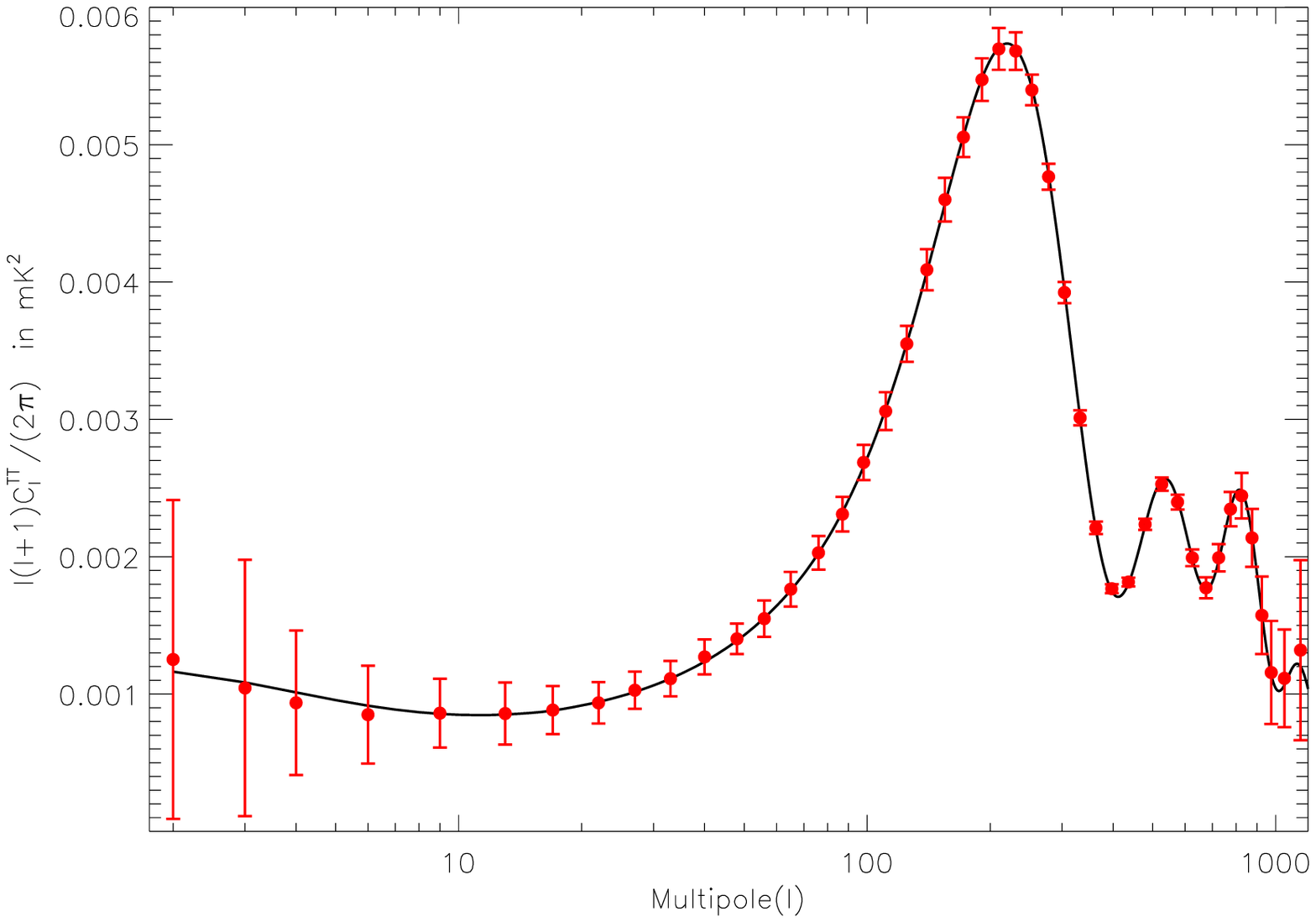}
  \caption{The red filled circles show the average estimated angular
    power spectrum using NILC from WMAP $7$-year-like simulations. The
    black solid line show the theoretical angular power spectrum for
    WMAP best-fit $\Lambda$CDM model. The top panel uses a linear
    scale in the horizontal axis, and the bottom panel a logarithmic
    one. }
  \label{fig:testcmb-cl}
\end{figure}

\subsubsection{Bias due to the method of spectrum estimation?}

Our method is based on the computation of the needlet transform of a
masked sky.  We check that the method itself is not biased, by
computing in the exact same way the power spectrum of 100 pure CMB
maps, randomly generated using different seeds. The average estimated
power spectrum is computed, and individual recovered power spectra for
the 100 realisations visually inspected. We find no evidence of a bias
in our power spectrum estimate.

\subsubsection{Bias due to the complete analysis chain?}

At a next level, we check whether our complete data processing method
can generate a bias. We generate two times 50 simulations of the NILC
maps for the 7 independent years. The first 50 coincide with the
simulations used to estimate an average ILC bias (green curve in
figure \ref{fig:cmb-cl-bias}) and correct for it in the real data (the
effect is small).  The other 50 simulations are used as test data, for
which we implement our complete analysis pipeline exactly as done on
the WMAP data.  For each of them, in particular, we de-bias the
resulting power spectrum using the average bias in the first 50
simulations, in the exact same way as was done on the real data.

Figure \ref{fig:testcmb-cl} shows the average angular power spectrum
(after bias correction) of CMB temperature anisotropies on these test
data sets.  The power spectrum matches extremely well the theoretical
input angular power spectrum, and does not show any lack of power at
large angular scale. This demonstrates that our method does not
generate biases by itself.  Nor the needlet ILC, nor the way we
implement our power spectrum estimation can be responsible for a bias
-- assuming our data model is right.

We note that the bias correction at $l=2$ is of more than $3\times
10^{-4}$ mK$^2$, and that the error on the estimate of this bias is
itself of order $5\times 10^{-4}$ mK$^2$. This can explain the
difference between the WMAP measured quadrupole, and our estimate.

\subsubsection{Are simulations representative?}

The representativeness of the simulated data sets is of course a major
concern.  For instance, if our simulated maps had exceedingly large
galactic foregrounds, the average bias on the simulations due to
residual foregrounds would be too large, and subtracting this residual
from the real data would result in an over-correction, and hence a
negative bias.

Although significant effort has been put to generate sky simulations
as realistic as possible, such a bias due to modelling errors cannot
be fully ruled-out (the modelling uncertainty is not easily
estimated).  For this reason, we look for confirmation using other CMB
maps and different analyses.

\subsubsection{Confirmation using other WMAP CMB maps?}

\begin{figure}
  \includegraphics[scale=0.42,angle=0]{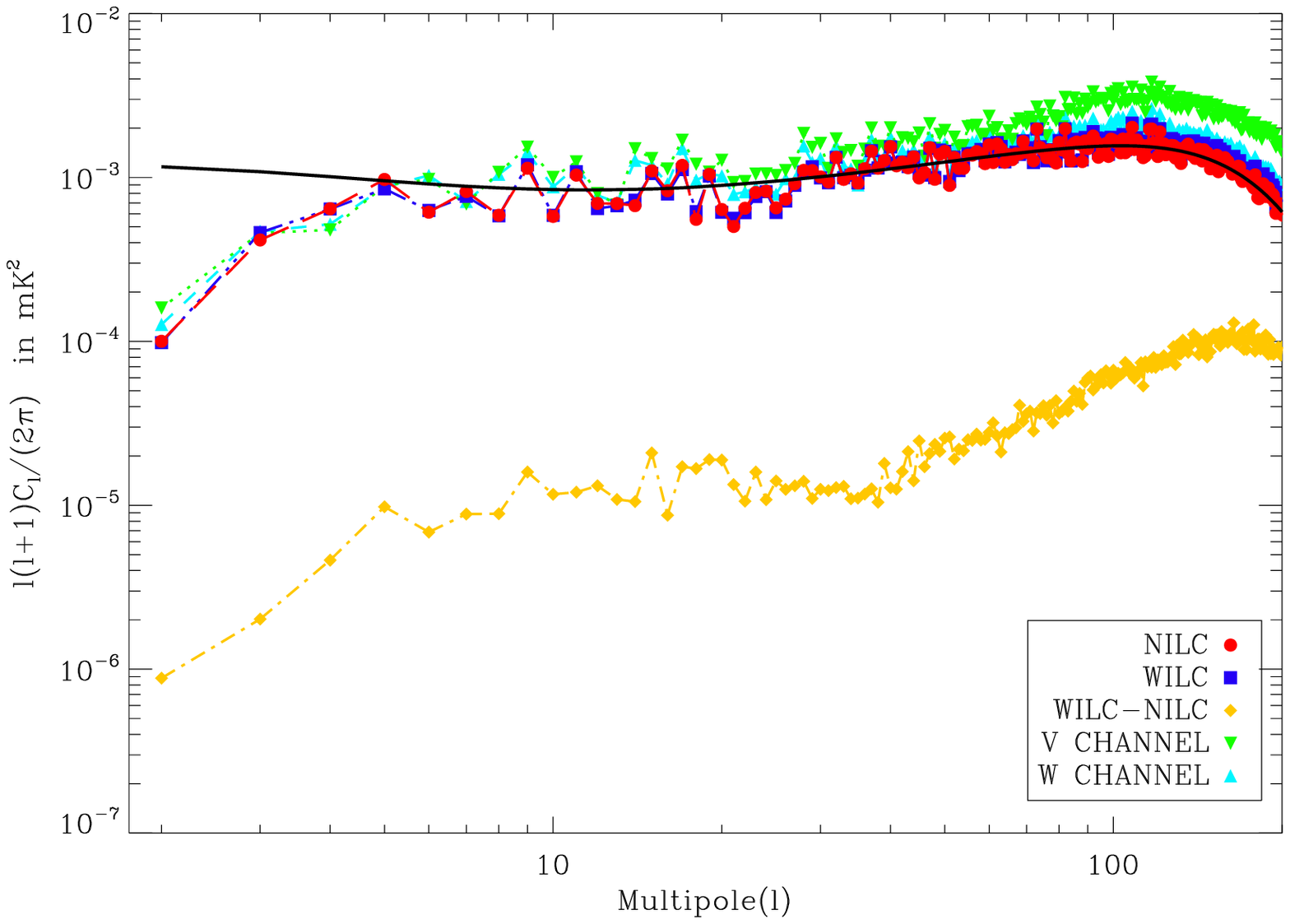}
  \includegraphics[scale=0.42,angle=0]{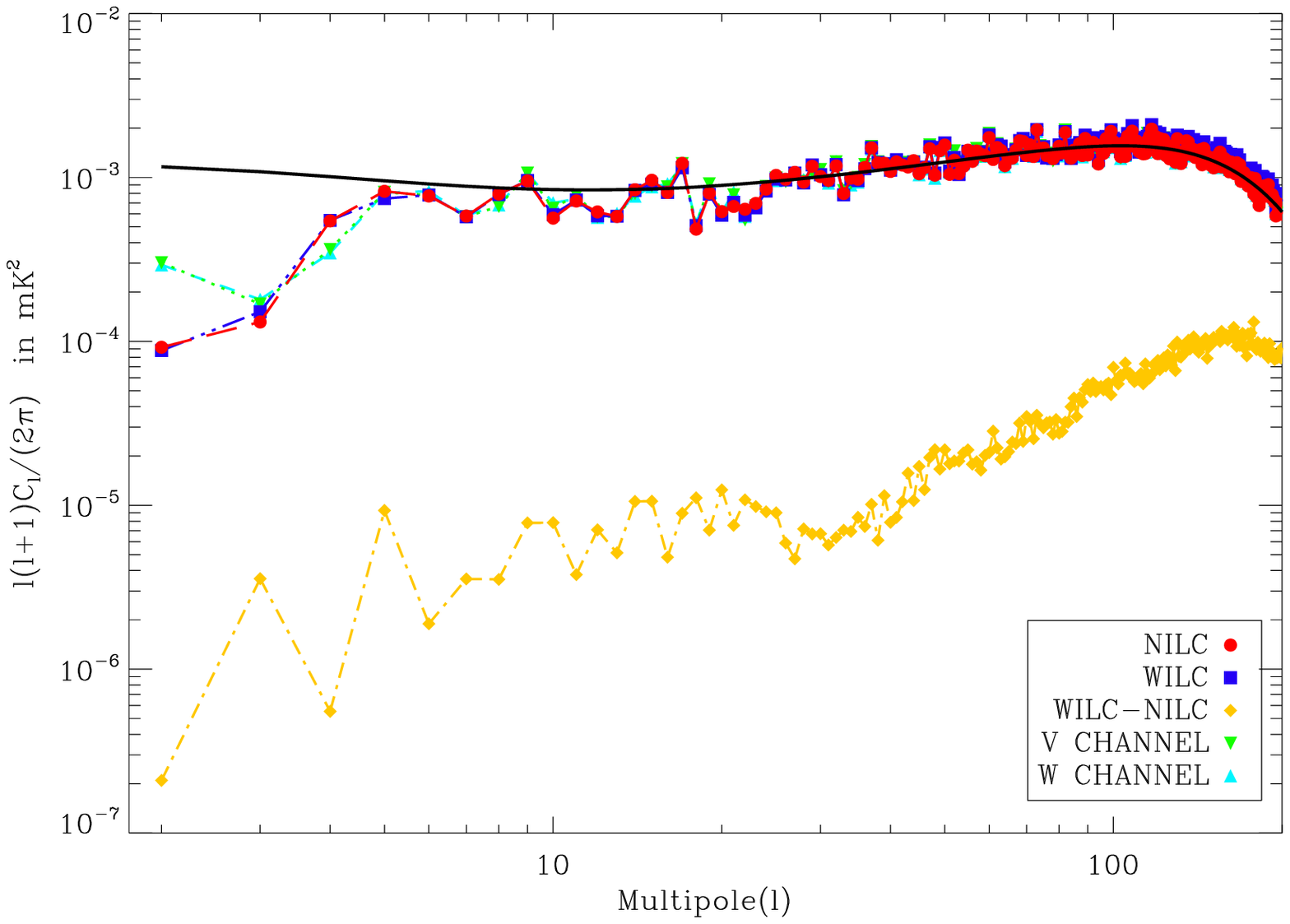}
  \includegraphics[scale=0.42,angle=0]{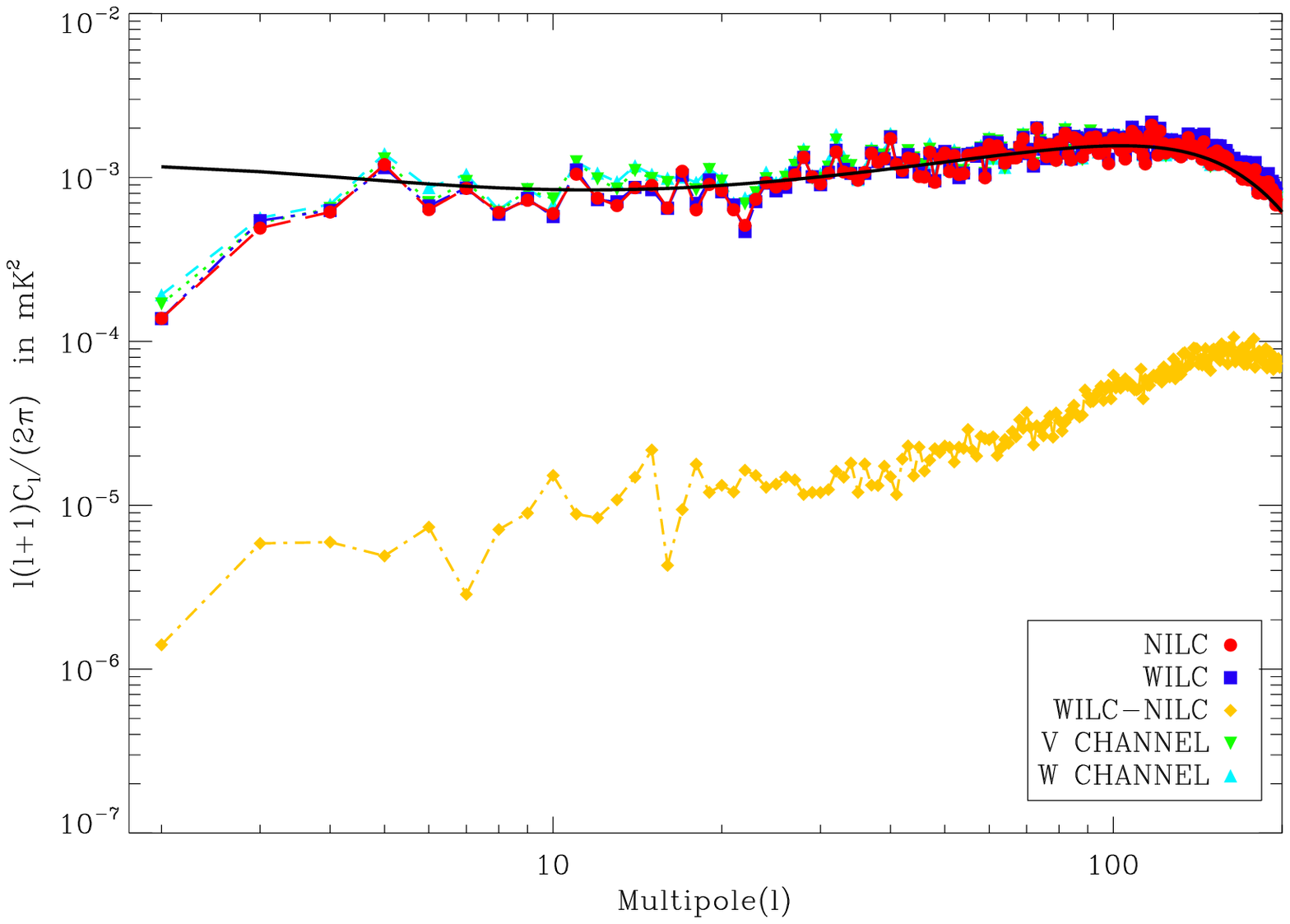}
  \caption{CMB power spectra obtained from different WMAP maps.  The
    red circles show the angular power spectrum obtained by direct
    calculation of the spectrum of the NILC CMB obtained on WMAP-7year
    band average maps, masked with an apodised Galactic mask. The mask
    used is $M_{10^{\circ},10^{\circ}}$ (same mask as used in our
    analysis) for the top panel, $M_{20^{\circ},20^{\circ}}$ for the
    middle panel, and kp2 (including point sources) for the bottom
    panel.  The 7-year NILC map power spectrum is fully consistent
    with the power spectrum of the WMAP 7-year ILC map (blue squares),
    masked in the same way. No correction for the ILC bias (which
    depends on the mask used) was applied here.  The yellow diamonds
    show the angular power spectrum of the difference of the two ILC
    maps.  This difference, very low on large scales, rises at higher
    $l$ because of the larger noise residuals in the WMAP ILC map. The
    downward and upward triangles show the power spectra of the 7-year
    V and W channels respectively.  As can be seen by comparing the
    various panels, galactic foregrounds contaminate significantly the
    V channel at galactic latitudes between 10 and 20 degrees (and, to
    a lesser extent, the W channel as well).  Note also that the
    lowest multipoles depend somewhat on the exact mask used, as
    expected for `pseudo' spectra.  }
  \label{fig:compare-cl}
\end{figure}

We have also estimated the power spectrum of WMAP maps directly using
the original V and W frequency channel maps, the foreground-reduced V
and W channel maps published by the WMAP collaboration, and the WMAP
ILC map itself, applying different galactic masks. These estimates
also result in low $\ell$ CMB multipoles.  In figure
\ref{fig:compare-cl} we compare the multipoles computed directly on
the 7-year NILC map and on the 7-year WMAP ILC. They are almost
indistinguishable at low $l$.  The very low value of the spectrum of
the difference between those two maps shows that the difference
between our present power spectrum and the published WMAP spectrum is
not due to significant differences in the large scale modes of the
maps themselves, but on the power spectrum estimation method
(including debiasing).

Power spectra computed directly from the 7-year V and W channels also
tend to be lower than the WMAP theoretical best fit. A fraction of the
deficit in power on the largest scales ($l=2$ to $l=4$) is due to the
mask (there is no bias correction here, except for a global $f_{\rm
  sky}$ factor). In most of the measured multipoles for $l<30$, there
seems to be less power in the maps than theoretically predicted. This
is particularly true for conservative masks
($M_{20^{\circ},20^{\circ}}$ and kp2).  We note that the power in the
V band map decreases with increasing galactic cut. For the
$M_{20^{\circ},20^{\circ}}$ mask, the spectra of all maps (Needlet
ILC, WMAP ILC, V band and W band) are almost indistinguishable, and on
average significantly lower than the theoretical best fit, for
$5<l<25$.

\subsection{Comparison with other work}

We have not been able to identify a source of bias in our analysis
which would explain this difference between our power spectrum and the
one published by the WMAP collaboration.  A significant amount of
connected work has been done by various authors to extract a CMB map,
and compute CMB power spectra from WMAP data, with different
techniques.  We now discuss how our present work differs from such
previous analyses with similar objectives, and compare the results
obtained with various methods and on various WMAP data releases (in
particular on large scales).

\citet{2009A&A...493..835D} make a CMB map from WMAP 5 year data on
needlet (spherical wavelet) domains. Our present work is an extension
implemented on WMAP $7$-year data. As an additional analysis step, we
also use the corresponding NILC weights to obtain clean CMB maps from
$7$ individual years, and compute cross spectra to estimate the CMB
power spectrum.

A CMB power spectrum has been published with each of the various WMAP
data releases
\citep{2003ApJS..148..135H,2007ApJS..170..288H,2009ApJS..180..296N,2011ApJS..192...16L}. The
CMB power spectrum on large scale is unexpectedly low in the analysis
of the first year data. In subsequent releases, only the quadrupole
remains significantly lower than the WMAP best fit model spectrum.

The most recent temperature power spectrum produced by the WMAP
collaboration is obtained on the basis of two different techniques for
large and small scales.  For $\ell \leq 32$, the spectrum is obtained
using a Blackwell-Rao estimator applied to a chain of Gibbs samples
based on the seven-year ILC map masked with the KQ85y7 mask, which
cuts 21.7~\% of the sky close to the galactic plane. The difference in
power spectrum at low multipoles between this estimate and ours must
originate from either the exact way the power spectrum is estimated on
large scales (including debiasing), but we do not know at present
whether the observed difference is to be expected on the basis of
differences in the analysis pipeline, nor its exact significance, as
this requires checking details of both analyses, some of which are not
available to us at present.

Among the differences, we note that our sky model is based on 4
galactic components (synchrotron, free-free, spinning dust and thermal
dust), with spectral indices varying over the sky for synchrotron and
thermal dust, while the WMAP 3-year analysis, for instance, uses a
three-component model of sky emission.  It is possible that a bias
correction based on simulations with a 3-component ISM emission yields
a higher $C_l$ estimate than with a 4-component one. If the origin of
the discrepancy is there, it emphasises the need for complete models
of sky emission, in which the complexity of the sky can be varied at
the level of present modelling uncertainties.

Another particularity of our analysis is the use of cross spectra
between data from different years of observations, which in principle
makes our analysis impervious to large scale residuals of low
frequency noise correlated between the various WMAP channels. Such
hypothetical residuals, however, are unlikely to be the main reason for the
low-$l$ discrepancy, as demonstrated by the comparison of the 7 year NILC map 
and the WMAP ILC map at low $l$.

At higher multipoles, the WMAP spectrum is estimated on the
seven-year, template-cleaned V- and W-band maps, using (pseudo)
cross-spectra between maps obtained for individual differencing
assemblies and individual years. Above $\ell=600$, the maps are
inverse-noise weighted to reduce the noise variance in the final power
spectrum estimate. The WMAP power spectrum is in good agreement with
our estimate.

In addition to the analyses published by the WMAP collaboration, the
CMB power spectrum inferred from WMAP data sets has been investigated
by a number of authors.

\citet{2007ApJ...656..641E}, in particular, partly re-analyse WMAP
3-year data sets with different methods.  They find large scale power
spectra systematically lower than the WMAP best fit and published
power spectrum, although not as low as our present result for $3\leq l
\leq 6$. All the estimated spectra they publish are somewhat
different, a feature they interpret, on the basis of differences when
the galactic mask is extended, as plausibly due to contamination by
residual galactic foreground emission.

On the foreground cleaning side, CMB reconstruction using ILC methods
has been performed by several authors. \citet{2008PhRvD..78b3003S}
have implemented an ILC on WMAP 1-year and 3-year data in the harmonic
domain. The primary goal of their work is to compute the CMB power
spectrum, rather than producing a clean CMB map. They also analyse the
bias produced at low $l$ by the ILC in the measured power
spectrum. This bias is estimated in the case of an implementation in
the harmonic domain, and an equivalent bias correction is made in our
case on the basis of Monte-Carlo simulations. In their analysis,
\citet{2008PhRvD..78b3003S} also find, as we do in our present
analysis, that both the quadrupole and the octupole are significantly
lower than the WMAP best fit.

 \citet{2010ApJ...714..840S} estimate CMB polarisation
and temperature power spectra using linear combination of WMAP 5 year
maps. The authors use different combinations of
individual WMAP differencing assemblies, rather than different years,
to compute independent sky maps, but find, again, somewhat lower CMB
power on large scale than both the WMAP estimate and best fit.

Hence, significant differences between the low CMB multipoles are
found by different authors, using different methods.  All the analyses
are based on the same sky emission, and hence should agree to up to
errors which exclude cosmic variance. Even worse, these analyses
sometimes start from the same original data, so that even the
instrumental noise errors should be the same.  This demonstrates that
the low $l$ power spectrum estimation is sensitive to details of the
analysis pipeline. The exact origin of the discrepancy between the
various estimates remains to be understood, but the fact that there
still is debate on the exact value of the low $l$ CMB multipoles is a
concern, which illustrates the difficulty of component separation and
power spectrum estimation on large scales. Solving these issues is
important for the tentative measurement of primordial tensor modes
with Planck and with future CMB polarisation experiments such as the
recently proposed CMBPol, EPIC, and COrE space missions
\citep{2009AIPC.1141....3B,2008arXiv0805.4207B,2011arXiv1102.2181T}.


\section{Conclusion}
\label{sec:conclusion}

The precise measurement of cosmological parameters has set a new goal
in the effort to estimate the angular power spectrum of the CMB by
present and future CMB experiments. This effort is primarily boosted
by the ever increasing improvement in the sensitivity and resolution
of the CMB observations. However, the removal of foreground
contamination from the observed sky maps is a mandatory preliminary
step for the accurate estimation of angular power spectrum.

In this paper, we have described a new methodology to estimate the
angular power spectrum of CMB temperature anisotropy from WMAP 7-year
data. We have used linear combination of sky maps decomposed on a
frame of spherical wavelets (needlets), to construct a map of CMB
anisotropies with low contamination from foreground signals and
instrumental noise.  The CMB temperature anisotropy map has been
estimated by implementing the NILC on WMAP $7$-year band average maps
at the highest resolution (with HEALPix pixelisation parameter nside
equal to $1024$). We have included, in our analysis, three foreground
templates in the set of analysed observations for better performance
of our component separation.  Our method for CMB cleaning does not
rely strongly upon any assumed model of foreground emission and
detector noise properties, and hence, is not very sensitive to the
uncertainty and insufficiency in foreground and noise
modelling. However, the cleaned CMB map always contains, at some
level, non-vanishing residual foreground and residual noise,
superimposed on actual CMB signal.

To minimise the impact of uncertainties in detector noise levels and
correlation between channels, the angular power spectrum of the CMB
has been estimated from all possible (twenty one in our case) cross
power spectra of clean CMB maps for $7$ individual years of
observations with independent noise. The estimates of CMB maps for the
$7$ individual years are obtained by using the NILC weights obtained
by implementing a needlet ILC on WMAP $7$-year band average maps.

To reduce the impact of foreground signals on the measured CMB power
spectrum, in addition to using low foreground linear combinations of
input WMAP channels and of ancillary data, we have applied the point
source mask provided by the WMAP collaboration, and then filled the
masked regions by a interpolation procedure.  We have also used an
apodised symmetric mask to lower the impact of residuals emission from
the interstellar medium in our analysis.

Biases due to random correlation of the CMB with the ISM in our NILC
pipeline have been estimated using $50$ realisations of WMAP-like
simulations.  These biases turn out to be smaller than the other
sources of error on most of the range of $l$, but not completely
negligible, so we use the average bias measured on these 50
simulations to de-bias our estimate on WMAP data.  We find that our
de-biased angular power spectrum agrees well with the estimate of
angular power spectrum provided by the WMAP collaboration, except at
large angular scale, where our estimate seems to be systematically
lower, but where the fundamental uncertainties are high. A number of
tests, performed on simulated data sets, using different approaches to
power spectrum estimation on WMAP data, confirm that the released WMAP
maps seems to contain somewhat less CMB anisotropies on large scales
than expected from the WMAP best fit model (marginally within cosmic
variance errors, as can be seen in the bottom panel of figure
\ref{fig:cmb-cl-diff}). This still has to be elucidated.

The error bars in our estimate of de-biased angular power spectrum
have been computed very carefully. They comprise estimates of the
measurement and cosmic variance error on the basis of the internal
scatter of 21 independent cross-spectra, and of the error in the
estimate of the ILC bias. Errors due to miscalibration or to imperfect
beam knowledge have been shown to be small enough to be negligible,
considering the calibration uncertainty estimates provided by the WMAP
collaboration. Our total error bars are comparable to the error in the
estimate of angular power spectrum provided by the WMAP
collaboration. As the measurement errors are estimated from the
internal scatter of independent cross spectra for each bin in $l$,
they do not rely on a model of WMAP detector noise.

The fact that there still is not a convincing consensus in the
scientific community on the exact value of the low $l$ CMB multipoles
is a concern, which demonstrates the difficulty of extracting
precisely the CMB emission from galactic foregrounds on large scales.


\section*{Acknowledgements}

Soumen Basak is supported by a `Physique des deux infinis' (P2I)
postdoctoral fellowship. We acknowledge the use of the Legacy Archive
for Microwave Background Data Analysis (LAMBDA). Support for LAMBDA is
provided by the NASA Office of Space Science.  The results in this
paper have been derived using the HEALPix package
\citep{2005ApJ...622..759G}.  The authors acknowledge the use of the
Planck Sky Model, developed by the Planck working group on component
separation, for making the simulations used in this work.  We thank
Jean-Fran\c{c}ois Cardoso, Guillaume Castex, Maude Le Jeune, Mathieu
Remazeilles and Radek Stompor for useful discussions.


\bibliography{wmap7yrTT}

\begin{thebibliography}{54}
\expandafter\ifx\csname natexlab\endcsname\relax\def\natexlab#1{#1}\fi

\bibitem[{{Atrio-Barandela} {et~al.}(2008){Atrio-Barandela}, {Kashlinsky},
  {Kocevski}, \& {Ebeling}}]{2008ApJ...675L..57A}
{Atrio-Barandela}, F., {Kashlinsky}, A., {Kocevski}, D., \& {Ebeling}, H. 2008,
  \apjl, 675, L57

\bibitem[{{Baumann} {et~al.}(2009){Baumann}, {Cooray}, {Dodelson}, {Dunkley},
  {Fraisse}, {Jackson}, {Kogut}, {Krauss}, {Smith}, \&
  {Zaldarriaga}}]{2009AIPC.1141....3B}
{Baumann}, D., {Cooray}, A., {Dodelson}, S., {Dunkley}, J., {Fraisse}, A.~A.,
  {Jackson}, M.~G., {Kogut}, A., {Krauss}, L.~M., {Smith}, K.~M., \&
  {Zaldarriaga}, M. 2009, in American Institute of Physics Conference Series,
  Vol. 1141, American Institute of Physics Conference Series, ed. {S.~Dodelson,
  D.~Baumann, A.~Cooray, J.~Dunkley, A.~Fraisse, M.~G.~Jackson, A.~Kogut,
  L.~Krauss, M.~Zaldarriaga, \& K.~Smith }, 3--9

\bibitem[{{Betoule} {et~al.}(2009){Betoule}, {Pierpaoli}, {Delabrouille}, {Le
  Jeune}, \& {Cardoso}}]{2009A&A...503..691B}
{Betoule}, M., {Pierpaoli}, E., {Delabrouille}, J., {Le Jeune}, M., \&
  {Cardoso}, J. 2009, \aap, 503, 691

\bibitem[{{Bock} {et~al.}(2008){Bock}, {Cooray}, {Hanany}, {Keating}, {Lee},
  {Matsumura}, {Milligan}, {Ponthieu}, {Renbarger}, \&
  {Tran}}]{2008arXiv0805.4207B}
{Bock}, J., {Cooray}, A., {Hanany}, S., {Keating}, B., {Lee}, A., {Matsumura},
  T., {Milligan}, M., {Ponthieu}, N., {Renbarger}, T., \& {Tran}, H. 2008,
  ArXiv e-prints

\bibitem[{{Bonaldi} {et~al.}(2007){Bonaldi}, {Ricciardi}, {Leach}, {Stivoli},
  {Baccigalupi}, \& {de Zotti}}]{2007MNRAS.382.1791B}
{Bonaldi}, A., {Ricciardi}, S., {Leach}, S., {Stivoli}, F., {Baccigalupi}, C.,
  \& {de Zotti}, G. 2007, \mnras, 382, 1791

\bibitem[{{Bottino} {et~al.}(2008){Bottino}, {Banday}, \&
  {Maino}}]{2008MNRAS.389.1190B}
{Bottino}, M., {Banday}, A.~J., \& {Maino}, D. 2008, \mnras, 389, 1190

\bibitem[{{Bottino} {et~al.}(2010){Bottino}, {Banday}, \&
  {Maino}}]{2010MNRAS.402..207B}
---. 2010, \mnras, 402, 207

\bibitem[{{Condon} {et~al.}(1998){Condon}, {Cotton}, {Greisen}, {Yin},
  {Perley}, {Taylor}, \& {Broderick}}]{1998AJ....115.1693C}
{Condon}, J.~J., {Cotton}, W.~D., {Greisen}, E.~W., {Yin}, Q.~F., {Perley},
  R.~A., {Taylor}, G.~B., \& {Broderick}, J.~J. 1998, \aj, 115, 1693

\bibitem[{{Delabrouille} \& {Cardoso}(2009)}]{2009LNP...665..159D}
{Delabrouille}, J. \& {Cardoso}, J. 2009, in Lecture Notes in Physics, Berlin
  Springer Verlag, Vol. 665, Data Analysis in Cosmology, ed.
  {V.~J.~Mart{\'{\i}}nez, E.~Saar, E.~Mart{\'{\i}}nez-Gonz{\'a}lez, \&
  M.-J.~Pons-Border{\'{\i}}a}, 159--205

\bibitem[{{Delabrouille} {et~al.}(2009){Delabrouille}, {Cardoso}, {Le Jeune},
  {Betoule}, {Fay}, \& {Guilloux}}]{2009A&A...493..835D}
{Delabrouille}, J., {Cardoso}, J.-F., {Le Jeune}, M., {Betoule}, M., {Fay}, G.,
  \& {Guilloux}, F. 2009, \aap, 493, 835

\bibitem[{{Dick} {et~al.}(2010){Dick}, {Remazeilles}, \&
  {Delabrouille}}]{2010MNRAS.401.1602D}
{Dick}, J., {Remazeilles}, M., \& {Delabrouille}, J. 2010, \mnras, 401, 1602

\bibitem[{{Eriksen} {et~al.}(2004){Eriksen}, {Banday}, {G{\'o}rski}, \&
  {Lilje}}]{2004ApJ...612..633E}
{Eriksen}, H.~K., {Banday}, A.~J., {G{\'o}rski}, K.~M., \& {Lilje}, P.~B. 2004,
  \apj, 612, 633

\bibitem[{{Eriksen} {et~al.}(2007){Eriksen}, {Huey}, {Saha}, {Hansen}, {Dick},
  {Banday}, {G{\'o}rski}, {Jain}, {Jewell}, {Knox}, {Larson}, {O'Dwyer},
  {Souradeep}, \& {Wandelt}}]{2007ApJ...656..641E}
{Eriksen}, H.~K., {Huey}, G., {Saha}, R., {Hansen}, F.~K., {Dick}, J.,
  {Banday}, A.~J., {G{\'o}rski}, K.~M., {Jain}, P., {Jewell}, J.~B., {Knox},
  L., {Larson}, D.~L., {O'Dwyer}, I.~J., {Souradeep}, T., \& {Wandelt}, B.~D.
  2007, \apj, 656, 641

\bibitem[{{Fa{\"y}} {et~al.}(2008){Fa{\"y}}, {Guilloux}, {Betoule}, {Cardoso},
  {Delabrouille}, \& {Le Jeune}}]{2008PhRvD..78h3013F}
{Fa{\"y}}, G., {Guilloux}, F., {Betoule}, M., {Cardoso}, J.-F., {Delabrouille},
  J., \& {Le Jeune}, M. 2008, \prd, 78, 083013

\bibitem[{{Finkbeiner}(2003)}]{2003ApJS..146..407F}
{Finkbeiner}, D.~P. 2003, \apjs, 146, 407

\bibitem[{{Gawiser} {et~al.}(1998){Gawiser}, {Finkbeiner}, {Jaffe}, {Baker},
  {Balbi}, {Davis}, {Hanany}, {Holzapfel}, {Moustakas}, {Robinson},
  {Scannapieco}, {Smoot}, \& {Silk}}]{1998astro.ph.12237G}
{Gawiser}, E., {Finkbeiner}, D., {Jaffe}, A., {Baker}, J.~C., {Balbi}, A.,
  {Davis}, M., {Hanany}, S., {Holzapfel}, W., {Moustakas}, L., {Robinson}, J.,
  {Scannapieco}, E., {Smoot}, G.~F., \& {Silk}, J. 1998, ArXiv Astrophysics
  e-prints

\bibitem[{{Ghosh} {et~al.}(2011){Ghosh}, {Delabrouille}, {Remazeilles},
  {Cardoso}, \& {Souradeep}}]{2011MNRAS.412..883G}
{Ghosh}, T., {Delabrouille}, J., {Remazeilles}, M., {Cardoso}, J.-F., \&
  {Souradeep}, T. 2011, \mnras, 412, 883

\bibitem[{{G{\'o}rski} {et~al.}(2005){G{\'o}rski}, {Hivon}, {Banday},
  {Wandelt}, {Hansen}, {Reinecke}, \& {Bartelmann}}]{2005ApJ...622..759G}
{G{\'o}rski}, K., {Hivon}, E., {Banday}, A., {Wandelt}, B., {Hansen}, F.,
  {Reinecke}, M., \& {Bartelmann}, M. 2005, Astrophys. J., 622, 759

\bibitem[{{Griffith} {et~al.}(1994){Griffith}, {Wright}, {Burke}, \&
  {Ekers}}]{1994ApJS...90..179G}
{Griffith}, M.~R., {Wright}, A.~E., {Burke}, B.~F., \& {Ekers}, R.~D. 1994,
  \apjs, 90, 179

\bibitem[{{Griffith} {et~al.}(1995){Griffith}, {Wright}, {Burke}, \&
  {Ekers}}]{1995ApJS...97..347G}
---. 1995, \apjs, 97, 347

\bibitem[{Guilloux {et~al.}(2009)Guilloux, Fa\"y, \&
  Cardoso}]{guilloux:fay:cardoso:2008}
Guilloux, F., Fa\"y, G., \& Cardoso, J.-F. 2009, Appl. Comput. Harmon. Anal.,
  26, 143

\bibitem[{{Haslam} {et~al.}(1981){Haslam}, {Klein}, {Salter}, {Stoffel},
  {Wilson}, {Cleary}, {Cooke}, \& {Thomasson}}]{1981A&A...100..209H}
{Haslam}, C.~G.~T., {Klein}, U., {Salter}, C.~J., {Stoffel}, H., {Wilson},
  W.~E., {Cleary}, M.~N., {Cooke}, D.~J., \& {Thomasson}, P. 1981, \aap, 100,
  209

\bibitem[{{Hinshaw} {et~al.}(2007){Hinshaw}, {Nolta}, {Bennett}, {Bean},
  {Dor{\'e}}, {Greason}, {Halpern}, {Hill}, {Jarosik}, {Kogut}, {Komatsu},
  {Limon}, {Odegard}, {Meyer}, {Page}, {Peiris}, {Spergel}, {Tucker}, {Verde},
  {Weiland}, {Wollack}, \& {Wright}}]{2007ApJS..170..288H}
{Hinshaw}, G., {Nolta}, M.~R., {Bennett}, C.~L., {Bean}, R., {Dor{\'e}}, O.,
  {Greason}, M.~R., {Halpern}, M., {Hill}, R.~S., {Jarosik}, N., {Kogut}, A.,
  {Komatsu}, E., {Limon}, M., {Odegard}, N., {Meyer}, S.~S., {Page}, L.,
  {Peiris}, H.~V., {Spergel}, D.~N., {Tucker}, G.~S., {Verde}, L., {Weiland},
  J.~L., {Wollack}, E., \& {Wright}, E.~L. 2007, \apjs, 170, 288

\bibitem[{{Hinshaw} {et~al.}(2003){Hinshaw}, {Spergel}, {Verde}, {Hill},
  {Meyer}, {Barnes}, {Bennett}, {Halpern}, {Jarosik}, {Kogut}, {Komatsu},
  {Limon}, {Page}, {Tucker}, {Weiland}, {Wollack}, \&
  {Wright}}]{2003ApJS..148..135H}
{Hinshaw}, G., {Spergel}, D.~N., {Verde}, L., {Hill}, R.~S., {Meyer}, S.~S.,
  {Barnes}, C., {Bennett}, C.~L., {Halpern}, M., {Jarosik}, N., {Kogut}, A.,
  {Komatsu}, E., {Limon}, M., {Page}, L., {Tucker}, G.~S., {Weiland}, J.~L.,
  {Wollack}, E., \& {Wright}, E.~L. 2003, \apjs, 148, 135

\bibitem[{{Jarosik} {et~al.}(2010){Jarosik}, {Bennett}, {Dunkley}, {Gold},
  {Greason}, {Halpern}, {Hill}, {Hinshaw}, {Kogut}, {Komatsu}, {Larson},
  {Limon}, {Meyer}, {Nolta}, {Odegard}, {Page}, {Smith}, {Spergel}, {Tucker},
  {Weiland}, {Wollack}, \& {Wright}}]{2010arXiv1001.4744J}
{Jarosik}, N., {Bennett}, C.~L., {Dunkley}, J., {Gold}, B., {Greason}, M.~R.,
  {Halpern}, M., {Hill}, R.~S., {Hinshaw}, G., {Kogut}, A., {Komatsu}, E.,
  {Larson}, D., {Limon}, M., {Meyer}, S.~S., {Nolta}, M.~R., {Odegard}, N.,
  {Page}, L., {Smith}, K.~M., {Spergel}, D.~N., {Tucker}, G.~S., {Weiland},
  J.~L., {Wollack}, E., \& {Wright}, E.~L. 2010, ArXiv e-prints

\bibitem[{{Kim} {et~al.}(2009){Kim}, {Naselsky}, \&
  {Christensen}}]{2009PhRvD..79b3003K}
{Kim}, J., {Naselsky}, P., \& {Christensen}, P.~R. 2009, \prd, 79, 023003

\bibitem[{{Komatsu} {et~al.}(2010){Komatsu}, {Smith}, {Dunkley}, {Bennett},
  {Gold}, {Hinshaw}, {Jarosik}, {Larson}, {Nolta}, {Page}, {Spergel},
  {Halpern}, {Hill}, {Kogut}, {Limon}, {Meyer}, {Odegard}, {Tucker}, {Weiland},
  {Wollack}, \& {Wright}}]{2010arXiv1001.4538K}
{Komatsu}, E., {Smith}, K.~M., {Dunkley}, J., {Bennett}, C.~L., {Gold}, B.,
  {Hinshaw}, G., {Jarosik}, N., {Larson}, D., {Nolta}, M.~R., {Page}, L.,
  {Spergel}, D.~N., {Halpern}, M., {Hill}, R.~S., {Kogut}, A., {Limon}, M.,
  {Meyer}, S.~S., {Odegard}, N., {Tucker}, G.~S., {Weiland}, J.~L., {Wollack},
  E., \& {Wright}, E.~L. 2010, ArXiv e-prints

\bibitem[{{Larson} {et~al.}(2011){Larson}, {Dunkley}, {Hinshaw}, {Komatsu},
  {Nolta}, {Bennett}, {Gold}, {Halpern}, {Hill}, {Jarosik}, {Kogut}, {Limon},
  {Meyer}, {Odegard}, {Page}, {Smith}, {Spergel}, {Tucker}, {Weiland},
  {Wollack}, \& {Wright}}]{2011ApJS..192...16L}
{Larson}, D., {Dunkley}, J., {Hinshaw}, G., {Komatsu}, E., {Nolta}, M.~R.,
  {Bennett}, C.~L., {Gold}, B., {Halpern}, M., {Hill}, R.~S., {Jarosik}, N.,
  {Kogut}, A., {Limon}, M., {Meyer}, S.~S., {Odegard}, N., {Page}, L., {Smith},
  K.~M., {Spergel}, D.~N., {Tucker}, G.~S., {Weiland}, J.~L., {Wollack}, E., \&
  {Wright}, E.~L. 2011, \apjs, 192, 16

\bibitem[{{Leach} {et~al.}(2008){Leach}, {Cardoso}, {Baccigalupi}, {Barreiro},
  {Betoule}, {Bobin}, {Bonaldi}, {Delabrouille}, {de Zotti}, {Dickinson},
  {Eriksen}, {Gonz{\'a}lez-Nuevo}, {Hansen}, {Herranz}, {Le Jeune},
  {L{\'o}pez-Caniego}, {Mart{\'{\i}}nez-Gonz{\'a}lez}, {Massardi}, {Melin},
  {Miville-Desch{\^e}nes}, {Patanchon}, {Prunet}, {Ricciardi}, {Salerno},
  {Sanz}, {Starck}, {Stivoli}, {Stolyarov}, {Stompor}, \&
  {Vielva}}]{2008A&A...491..597L}
{Leach}, S.~M., {Cardoso}, J.-F., {Baccigalupi}, C., {Barreiro}, R.~B.,
  {Betoule}, M., {Bobin}, J., {Bonaldi}, A., {Delabrouille}, J., {de Zotti},
  G., {Dickinson}, C., {Eriksen}, H.~K., {Gonz{\'a}lez-Nuevo}, J., {Hansen},
  F.~K., {Herranz}, D., {Le Jeune}, M., {L{\'o}pez-Caniego}, M.,
  {Mart{\'{\i}}nez-Gonz{\'a}lez}, E., {Massardi}, M., {Melin}, J.,
  {Miville-Desch{\^e}nes}, M., {Patanchon}, G., {Prunet}, S., {Ricciardi}, S.,
  {Salerno}, E., {Sanz}, J.~L., {Starck}, J., {Stivoli}, F., {Stolyarov}, V.,
  {Stompor}, R., \& {Vielva}, P. 2008, \aap, 491, 597

\bibitem[{{Maino} {et~al.}(2007){Maino}, {Donzelli}, {Banday}, {Stivoli}, \&
  {Baccigalupi}}]{2007MNRAS.374.1207M}
{Maino}, D., {Donzelli}, S., {Banday}, A.~J., {Stivoli}, F., \& {Baccigalupi},
  C. 2007, \mnras, 374, 1207

\bibitem[{{Marinucci} {et~al.}(2008){Marinucci}, {Pietrobon}, {Balbi}, {Baldi},
  {Cabella}, {Kerkyacharian}, {Natoli}, {Picard}, \&
  {Vittorio}}]{2008MNRAS.383..539M}
{Marinucci}, D., {Pietrobon}, D., {Balbi}, A., {Baldi}, P., {Cabella}, P.,
  {Kerkyacharian}, G., {Natoli}, P., {Picard}, D., \& {Vittorio}, N. 2008,
  \mnras, 383, 539

\bibitem[{{Mauch} {et~al.}(2003){Mauch}, {Murphy}, {Buttery}, {Curran},
  {Hunstead}, {Piestrzynski}, {Robertson}, \& {Sadler}}]{2003MNRAS.342.1117M}
{Mauch}, T., {Murphy}, T., {Buttery}, H.~J., {Curran}, J., {Hunstead}, R.~W.,
  {Piestrzynski}, B., {Robertson}, J.~G., \& {Sadler}, E.~M. 2003, \mnras, 342,
  1117

\bibitem[{{Melin} {et~al.}(2011){Melin}, {Bartlett}, {Delabrouille}, {Arnaud},
  {Piffaretti}, \& {Pratt}}]{2011A&A...525A.139M}
{Melin}, J.-B., {Bartlett}, J.~G., {Delabrouille}, J., {Arnaud}, M.,
  {Piffaretti}, R., \& {Pratt}, G.~W. 2011, \aap, 525, A139+

\bibitem[{{Miville-Desch{\^e}nes} {et~al.}(2008){Miville-Desch{\^e}nes},
  {Ysard}, {Lavabre}, {Ponthieu}, {Mac{\'{\i}}as-P{\'e}rez}, {Aumont}, \&
  {Bernard}}]{2008A&A...490.1093M}
{Miville-Desch{\^e}nes}, M.-A., {Ysard}, N., {Lavabre}, A., {Ponthieu}, N.,
  {Mac{\'{\i}}as-P{\'e}rez}, J.~F., {Aumont}, J., \& {Bernard}, J.~P. 2008,
  \aap, 490, 1093

\bibitem[{{Narcowich} {et~al.}(2006){Narcowich}, {Petrushev}, \&
  {Ward}}]{narcowich:petrushev:ward:2006}
{Narcowich}, F., {Petrushev}, P., \& {Ward}, J. 2006, SIAM J. Math. Anal., 38,
  574

\bibitem[{{Nolta} {et~al.}(2009){Nolta}, {Dunkley}, {Hill}, {Hinshaw},
  {Komatsu}, {Larson}, {Page}, {Spergel}, {Bennett}, {Gold}, {Jarosik},
  {Odegard}, {Weiland}, {Wollack}, {Halpern}, {Kogut}, {Limon}, {Meyer},
  {Tucker}, \& {Wright}}]{2009ApJS..180..296N}
{Nolta}, M.~R., {Dunkley}, J., {Hill}, R.~S., {Hinshaw}, G., {Komatsu}, E.,
  {Larson}, D., {Page}, L., {Spergel}, D.~N., {Bennett}, C.~L., {Gold}, B.,
  {Jarosik}, N., {Odegard}, N., {Weiland}, J.~L., {Wollack}, E., {Halpern}, M.,
  {Kogut}, A., {Limon}, M., {Meyer}, S.~S., {Tucker}, G.~S., \& {Wright}, E.~L.
  2009, \apjs, 180, 296

\bibitem[{{Park} {et~al.}(2007){Park}, {Park}, \& {Gott}}]{2007ApJ...660..959P}
{Park}, C.-G., {Park}, C., \& {Gott}, J.~R.~I. 2007, \apj, 660, 959

\bibitem[{{Patanchon} {et~al.}(2005){Patanchon}, {Cardoso}, {Delabrouille}, \&
  {Vielva}}]{2005MNRAS.364.1185P}
{Patanchon}, G., {Cardoso}, J.-F., {Delabrouille}, J., \& {Vielva}, P. 2005,
  \mnras, 364, 1185

\bibitem[{{Penzias} \& {Wilson}(1965)}]{1965ApJ...142..419P}
{Penzias}, A.~A. \& {Wilson}, R.~W. 1965, \apj, 142, 419

\bibitem[{{Pietrobon} {et~al.}(2008){Pietrobon}, {Amblard}, {Balbi}, {Cabella},
  {Cooray}, \& {Marinucci}}]{2008PhRvD..78j3504P}
{Pietrobon}, D., {Amblard}, A., {Balbi}, A., {Cabella}, P., {Cooray}, A., \&
  {Marinucci}, D. 2008, \prd, 78, 103504

\bibitem[{{Planck Collaboration} {et~al.}(2011){Planck Collaboration}, {Ade},
  {Aghanim}, {Arnaud}, {Ashdown}, {Aumont}, {Baccigalupi}, {Baker}, {Balbi},
  {Banday}, \& et~al.}]{2011arXiv1101.2022P}
{Planck Collaboration}, {Ade}, P.~A.~R., {Aghanim}, N., {Arnaud}, M.,
  {Ashdown}, M., {Aumont}, J., {Baccigalupi}, C., {Baker}, M., {Balbi}, A.,
  {Banday}, A.~J., \& et~al. 2011, ArXiv e-prints

\bibitem[{{Remazeilles} {et~al.}(2011{\natexlab{a}}){Remazeilles},
  {Delabrouille}, \& {Cardoso}}]{2011MNRAS.410.2481R}
{Remazeilles}, M., {Delabrouille}, J., \& {Cardoso}, J.-F. 2011{\natexlab{a}},
  \mnras, 410, 2481

\bibitem[{{Remazeilles} {et~al.}(2011{\natexlab{b}}){Remazeilles},
  {Delabrouille}, \& {Cardoso}}]{2011arXiv1103.1166R}
---. 2011{\natexlab{b}}, ArXiv e-prints

\bibitem[{{Rudjord} {et~al.}(2009){Rudjord}, {Hansen}, {Lan}, {Liguori},
  {Marinucci}, \& {Matarrese}}]{2009ApJ...701..369R}
{Rudjord}, {\O}., {Hansen}, F.~K., {Lan}, X., {Liguori}, M., {Marinucci}, D.,
  \& {Matarrese}, S. 2009, \apj, 701, 369

\bibitem[{{Saha} {et~al.}(2008){Saha}, {Prunet}, {Jain}, \&
  {Souradeep}}]{2008PhRvD..78b3003S}
{Saha}, R., {Prunet}, S., {Jain}, P., \& {Souradeep}, T. 2008, \prd, 78, 023003

\bibitem[{{Samal} {et~al.}(2010){Samal}, {Saha}, {Delabrouille}, {Prunet},
  {Jain}, \& {Souradeep}}]{2010ApJ...714..840S}
{Samal}, P.~K., {Saha}, R., {Delabrouille}, J., {Prunet}, S., {Jain}, P., \&
  {Souradeep}, T. 2010, \apj, 714, 840

\bibitem[{{Schlegel} {et~al.}(1998){Schlegel}, {Finkbeiner}, \&
  {Davis}}]{1998ApJ...500..525S}
{Schlegel}, D.~J., {Finkbeiner}, D.~P., \& {Davis}, M. 1998, \apj, 500, 525

\bibitem[{{Souradeep}(2011)}]{2011BASI...39..163S}
{Souradeep}, T. 2011, Bulletin of the Astronomical Society of India, 39, 163

\bibitem[{{Tauber} {et~al.}(2010){Tauber}, {Mandolesi}, {Puget}, {Banos},
  {Bersanelli}, {Bouchet}, {Butler}, {Charra}, {Crone}, {Dodsworth}, \&
  et~al.}]{2010A&A...520A...1T}
{Tauber}, J.~A., {Mandolesi}, N., {Puget}, J., {Banos}, T., {Bersanelli}, M.,
  {Bouchet}, F.~R., {Butler}, R.~C., {Charra}, J., {Crone}, G., {Dodsworth},
  J., \& et~al. 2010, \aap, 520, A1+

\bibitem[{{Tegmark} {et~al.}(2003){Tegmark}, {de Oliveira-Costa}, \&
  {Hamilton}}]{2003PhRvD..68l3523T}
{Tegmark}, M., {de Oliveira-Costa}, A., \& {Hamilton}, A.~J. 2003, \prd, 68,
  123523

\bibitem[{{Tegmark} \& {Efstathiou}(1996)}]{1996MNRAS.281.1297T}
{Tegmark}, M. \& {Efstathiou}, G. 1996, \mnras, 281, 1297

\bibitem[{{The COrE Collaboration} {et~al.}(2011){The COrE Collaboration},
  {Armitage-Caplan}, {Avillez}, {Barbosa}, {Banday}, {Bartolo}, {Battye},
  {Bernard}, {de Bernardis}, {Basak}, {Bersanelli}, {Bielewicz}, {Bonaldi},
  {Bucher}, {Bouchet}, {Boulanger}, {Burigana}, {Camus}, {Challinor},
  {Chongchitnan}, {Clements}, {Colafrancesco}, {Delabrouille}, {De Petris}, {De
  Zotti}, {Dickinson}, {Dunkley}, {Ensslin}, {Fergusson}, {Ferreira},
  {Ferriere}, {Finelli}, {Galli}, {Garcia-Bellido}, {Gauthier}, {Haverkorn},
  {Hindmarsh}, {Jaffe}, {Kunz}, {Lesgourgues}, {Liddle}, {Liguori},
  {Lopez-Caniego}, {Maffei}, {Marchegiani}, {Martinez-Gonzalez}, {Masi},
  {Mauskopf}, {Matarrese}, {Melchiorri}, {Mukherjee}, {Nati}, {Natoli},
  {Negrello}, {Pagano}, {Paoletti}, {Peacocke}, {Peiris}, {Perroto},
  {Piacentini}, {Piat}, {Piccirillo}, {Pisano}, {Ponthieu}, {Rath},
  {Ricciardi}, {Rubino Martin}, {Salatino}, {Shellard}, {Stompor},
  {Urrestilla}, {Van Tent}, {Verde}, {Wandelt}, \&
  {Withington}}]{2011arXiv1102.2181T}
{The COrE Collaboration}, {Armitage-Caplan}, C., {Avillez}, M., {Barbosa}, D.,
  {Banday}, A., {Bartolo}, N., {Battye}, R., {Bernard}, J., {de Bernardis}, P.,
  {Basak}, S., {Bersanelli}, M., {Bielewicz}, P., {Bonaldi}, A., {Bucher}, M.,
  {Bouchet}, F., {Boulanger}, F., {Burigana}, C., {Camus}, P., {Challinor}, A.,
  {Chongchitnan}, S., {Clements}, D., {Colafrancesco}, S., {Delabrouille}, J.,
  {De Petris}, M., {De Zotti}, G., {Dickinson}, C., {Dunkley}, J., {Ensslin},
  T., {Fergusson}, J., {Ferreira}, P., {Ferriere}, K., {Finelli}, F., {Galli},
  S., {Garcia-Bellido}, J., {Gauthier}, C., {Haverkorn}, M., {Hindmarsh}, M.,
  {Jaffe}, A., {Kunz}, M., {Lesgourgues}, J., {Liddle}, A., {Liguori}, M.,
  {Lopez-Caniego}, M., {Maffei}, B., {Marchegiani}, P., {Martinez-Gonzalez},
  E., {Masi}, S., {Mauskopf}, P., {Matarrese}, S., {Melchiorri}, A.,
  {Mukherjee}, P., {Nati}, F., {Natoli}, P., {Negrello}, M., {Pagano}, L.,
  {Paoletti}, D., {Peacocke}, T., {Peiris}, H., {Perroto}, L., {Piacentini},
  F., {Piat}, M., {Piccirillo}, L., {Pisano}, G., {Ponthieu}, N., {Rath}, C.,
  {Ricciardi}, S., {Rubino Martin}, J., {Salatino}, M., {Shellard}, P.,
  {Stompor}, R., {Urrestilla}, L.~T.~J., {Van Tent}, B., {Verde}, L.,
  {Wandelt}, B., \& {Withington}, S. 2011, ArXiv e-prints

\bibitem[{{Wright} {et~al.}(1994){Wright}, {Griffith}, {Burke}, \&
  {Ekers}}]{1994ApJS...91..111W}
{Wright}, A.~E., {Griffith}, M.~R., {Burke}, B.~F., \& {Ekers}, R.~D. 1994,
  \apjs, 91, 111

\bibitem[{{Wright} {et~al.}(1996){Wright}, {Griffith}, {Hunt}, {Troup},
  {Burke}, \& {Ekers}}]{1996ApJS..103..145W}
{Wright}, A.~E., {Griffith}, M.~R., {Hunt}, A.~J., {Troup}, E., {Burke}, B.~F.,
  \& {Ekers}, R.~D. 1996, \apjs, 103, 145

\end{thebibliography}
\label{lastpage}

\end{document}